\documentclass[preprint,12pt]{elsarticle}
\biboptions{sort&compress}

\usepackage{graphicx}
\usepackage{amsmath}
\usepackage{amssymb}
\usepackage{bm}
\usepackage{bbm}
\usepackage{tikz}
\usepackage{subcaption}
\usepackage{hyperref}
\usepackage{relsize}
\usepackage{booktabs}
\usepackage{makecell}
\usepackage{siunitx}
\usepackage{enumitem}
\usepackage{xurl}
\usepackage{xcolor,graphicx,amssymb,tikz}
\usetikzlibrary{arrows.meta}
\usepackage{tcolorbox}
\tcbuselibrary{skins,breakable}
\usepackage{xurl}
\usepackage{iftex}
\ifPDFTeX
  \usepackage[scale=0.95]{tgheros}
  \newcommand{\chatmono}{\ttfamily}
\else
  \usepackage{fontspec}
  \newfontfamily\chatmono{lmmono10-regular.otf}[BoldFont=lmmonolt10-bold.otf]
\fi
\providecommand{\chatimgdir}{chat_transcript/}

\definecolor{uiPrimary}{HTML}{006DA3}
\definecolor{uiAccent}{HTML}{EFF3F4}
\definecolor{uiText}{HTML}{334155}
\definecolor{uiTextLight}{HTML}{64748B}
\definecolor{uiTextMuted}{HTML}{94A3B8}
\definecolor{uiSep}{HTML}{CBD5E1}
\definecolor{uiBorder}{HTML}{E2E8F0}
\definecolor{uiBgSubtle}{HTML}{F8FAFC}
\definecolor{uiHeading}{HTML}{1E293B}
\definecolor{uiSuccess}{HTML}{16A34A}
\definecolor{uiWarning}{HTML}{D97706}
\definecolor{uiError}{HTML}{DC2626}
\definecolor{uiViolet}{HTML}{7C3AED}
\definecolor{uiBlue}{HTML}{0284C7}
\definecolor{bSkyBg}{HTML}{E0F2FE}\definecolor{bSkyFg}{HTML}{0369A1}\definecolor{bSkyBd}{HTML}{7DD3FC}\definecolor{bSkyDot}{HTML}{0284C7}
\definecolor{bVioBg}{HTML}{EDE9FE}\definecolor{bVioFg}{HTML}{6D28D9}\definecolor{bVioBd}{HTML}{C4B5FD}\definecolor{bVioDot}{HTML}{7C3AED}
\definecolor{bGrnBg}{HTML}{ECFDF5}\definecolor{bGrnFg}{HTML}{047857}\definecolor{bGrnBd}{HTML}{6EE7B7}\definecolor{bGrnDot}{HTML}{059669}
\definecolor{bAmbBg}{HTML}{FFFBEB}\definecolor{bAmbFg}{HTML}{B45309}\definecolor{bAmbBd}{HTML}{FCD34D}\definecolor{bAmbDot}{HTML}{D97706}
\definecolor{bRosBg}{HTML}{FEF2F2}\definecolor{bRosFg}{HTML}{BE123C}\definecolor{bRosBd}{HTML}{FCA5A5}\definecolor{bRosDot}{HTML}{DC2626}
\definecolor{bBluBg}{HTML}{EFF6FF}\definecolor{bBluFg}{HTML}{1D4ED8}\definecolor{bBluBd}{HTML}{93C5FD}\definecolor{bBluDot}{HTML}{2563EB}
\definecolor{bCyaBg}{HTML}{ECFEFF}\definecolor{bCyaFg}{HTML}{0E7490}\definecolor{bCyaBd}{HTML}{67E8F9}\definecolor{bCyaDot}{HTML}{0891B2}
\definecolor{bOrgBg}{HTML}{FFF7ED}\definecolor{bOrgFg}{HTML}{C2410C}\definecolor{bOrgBd}{HTML}{FDBA74}\definecolor{bOrgDot}{HTML}{EA580C}
\definecolor{bIndBg}{HTML}{EEF2FF}\definecolor{bIndFg}{HTML}{4338CA}\definecolor{bIndBd}{HTML}{A5B4FC}\definecolor{bIndDot}{HTML}{4F46E5}
\definecolor{bMagBg}{HTML}{FDF4FF}\definecolor{bMagFg}{HTML}{A21CAF}\definecolor{bMagBd}{HTML}{E879F9}\definecolor{bMagDot}{HTML}{C026D3}

\newcommand{\chatmsgfont}{\sffamily\footnotesize}
\newcommand{\rowfont}{\sffamily\scriptsize}
\newcommand{\detfont}{\sffamily\fontsize{7}{8.4}\selectfont}
\newcommand{\codefont}{\chatmono\fontsize{7}{8.4}\selectfont}
\newcommand{\badgefont}{\sffamily\fontsize{6.3}{7.3}\selectfont\bfseries}
\newcommand{\hdrfont}{\sffamily\footnotesize\bfseries}
\newcommand{\btnfont}{\sffamily\scriptsize}
\newcommand{\namefont}{\sffamily\scriptsize\bfseries}

\newcommand{\iconcheck}[1]{\tikz[baseline=-0.6ex]{\draw[#1,line width=0.6pt] (0,0) circle (2.8pt);\draw[#1,line width=0.8pt,line cap=round,line join=round] (-1.4pt,0.1pt) -- (-0.45pt,-1.05pt) -- (1.5pt,1.05pt);}}
\newcommand{\iconwarn}[1]{\tikz[baseline=-0.6ex]{\draw[#1,line width=0.6pt,line join=round] (-3pt,-2.4pt) -- (3pt,-2.4pt) -- (0,2.9pt) -- cycle;\draw[#1,line width=0.7pt,line cap=round] (0,1.1pt) -- (0,-0.5pt);\fill[#1] (0,-1.5pt) circle (0.45pt);}}
\newcommand{\iconx}[1]{\tikz[baseline=-0.6ex]{\draw[#1,line width=0.6pt] (0,0) circle (2.8pt);\draw[#1,line width=0.7pt,line cap=round] (-1.2pt,-1.2pt) -- (1.2pt,1.2pt) (-1.2pt,1.2pt) -- (1.2pt,-1.2pt);}}
\newcommand{\iconmsg}[1]{\tikz[baseline=-0.6ex]{\draw[#1,line width=0.6pt,line join=round,rounded corners=0.5pt] (-2.7pt,2.1pt) -- (2.7pt,2.1pt) -- (2.7pt,-1.2pt) -- (-1pt,-1.2pt) -- (-2.7pt,-2.8pt) -- cycle;}}
\newcommand{\iconarrow}{\tikz[baseline=-0.6ex]{\draw[uiViolet,line width=0.7pt,-{Straight Barb[length=2pt,width=2.6pt]}] (0,0) -- (5pt,0);}}
\newcommand{\chevright}{\tikz[baseline=-0.5ex]{\draw[uiTextLight,line width=0.7pt,line cap=round,line join=round] (0,1.9pt) -- (1.7pt,0) -- (0,-1.9pt);}}
\newcommand{\chevdown}{\tikz[baseline=-0.5ex]{\draw[uiTextLight,line width=0.7pt,line cap=round,line join=round] (-1.9pt,0.9pt) -- (0,-0.9pt) -- (1.9pt,0.9pt);}}
\newcommand{\nodedot}[1]{\tikz[baseline=-0.5ex]\fill[b#1Dot] (0,0) circle (1.6pt);}

\newcommand{\agentavatar}{\begin{tikzpicture}[baseline=-2.6pt]\clip (0,0) circle (6pt);\node[inner sep=0pt] at (0,0){\includegraphics[width=12pt]{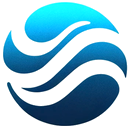}};\end{tikzpicture}}
\newcommand{\useravatar}{\begin{tikzpicture}[baseline=-2.6pt]\fill[uiTextMuted] (0,0) circle (6pt);\clip (0,0) circle (6pt);\fill[white] (0,1.7pt) circle (2.1pt);\fill[white] (0,-4.6pt) circle (4.3pt);\end{tikzpicture}}
\newcommand{\agentturnhead}{\par\vspace{7pt}\noindent\agentavatar\hspace{5pt}{\namefont\color{uiText}Agent}\par\nobreak\vspace{2pt}}
\newcommand{\userturnhead}{\par\vspace{7pt}\noindent\hfill{\namefont\color{uiText}User}\hspace{5pt}\useravatar\par\nobreak\vspace{2pt}}

\usepackage{eso-pic,refcount}
\usetikzlibrary{calc}
\newif\ifchatframe
\AddToShipoutPictureBG{\ifchatframe\AtTextLowerLeft{%
  \begin{tikzpicture}[remember picture,overlay]
    \ifnum\value{page}=\getpagerefnumber{chatframe:start}\relax
      \draw[uiSep,line width=0.6pt,rounded corners=5pt]
        let \p1=(chatframe start) in (-9pt,-9pt) rectangle (\textwidth+9pt,\y1);
    \else
      \draw[uiSep,line width=0.6pt,rounded corners=5pt] (-9pt,-9pt) rectangle (\textwidth+9pt,\textheight+9pt);
    \fi
  \end{tikzpicture}}\fi}
\newcommand{\chatlineskip}{\lineskip=1pt\lineskiplimit=0pt}
\newcommand{\chatframeon}{\chatlineskip\par\begingroup\dimen0=\pagegoal\advance\dimen0 by -\pagetotal\ifdim\dimen0<8\baselineskip\newpage\fi\endgroup\vspace{9pt}\noindent
  \tikz[remember picture,overlay]\coordinate (chatframe start) at (0,9pt);\label{chatframe:start}%
  \chatframetrue\par\vspace{-\baselineskip}}
\newcommand{\chatframeoff}{\clearpage\chatframefalse}

\newenvironment{agentbody}{\begin{list}{}{\leftmargin=17pt\rightmargin=0pt\topsep=0pt\parsep=0pt\itemsep=0pt\partopsep=0pt\listparindent=0pt\itemindent=0pt\labelwidth=0pt\labelsep=0pt}\item\relax\chatmsgfont\color{uiHeading}\rightskip=0pt plus 1fil\relax}{\end{list}\par\vspace{3pt}}
\newenvironment{agentmsg}{\agentturnhead\begin{agentbody}}{\end{agentbody}}
\newenvironment{agentmsgc}{\par\vspace{4pt}\begin{agentbody}}{\end{agentbody}}
\newtcolorbox{usermsg}{enhanced,breakable,colback=uiAccent,colframe=uiAccent,boxrule=0pt,arc=8pt,
  width=0.8\linewidth,flush right,halign=left,left=9pt,right=9pt,top=5pt,bottom=5pt,boxsep=0pt,
  before skip=0pt,after skip=3pt,fontupper=\chatmsgfont\color{uiHeading},before={\userturnhead}}

\newcommand{\steprow}[1]{\par\vspace{1.5pt}\noindent\chevright\hspace{4pt}{\rowfont\color{uiTextLight}Used #1}\par\vspace{1.5pt}}
\newcommand{\badge}[2][Sky]{\tcbox[on line,enhanced,colback=b#1Bg,colframe=b#1Bd,boxrule=0.4pt,arc=1.3pt,
  left=2.5pt,right=2.5pt,top=0.2pt,bottom=0.2pt,boxsep=0.5pt,fontupper=\badgefont\color{b#1Fg}]{#2}}
\newcommand{\stageheader}[3][Sky]{\par\vspace{4pt}\noindent\nodedot{#1}\hspace{4pt}{\hdrfont\color{uiHeading}#2}\hspace{4pt}{\color{uiSep}$\cdot$}\hspace{4pt}{\rowfont\color{uiTextLight}#3}\par\vspace{-1.5pt}\noindent{\color{uiBorder}\rule{\linewidth}{0.4pt}}\par\vspace{1.5pt}}
\newcommand{\inforow}[3][Sky]{\par\noindent\hangindent=10pt\hangafter=1\badge[#1]{#2}\hspace{4pt}{\rowfont\color{uiText}#3}\par}
\newcommand{\okrow}[1]{\par\noindent\hangindent=11pt\hangafter=1\iconcheck{uiSuccess}\hspace{3.5pt}{\rowfont\color{uiSuccess}#1}\par}
\newcommand{\warnrow}[1]{\par\noindent\hangindent=11pt\hangafter=1\iconwarn{uiWarning}\hspace{3.5pt}{\rowfont\color{uiWarning}#1}\par}
\newcommand{\errrow}[1]{\par\noindent\hangindent=11pt\hangafter=1\iconx{uiError}\hspace{3.5pt}{\rowfont\color{uiError}#1}\par}
\newcommand{\detailrow}[1]{\par\noindent\hangindent=10pt\hangafter=0\hspace*{10pt}{\detfont\color{uiTextMuted}#1}\par}
\newcommand{\kvrow}[2]{\par\noindent\hangindent=10pt\hangafter=0\hspace*{10pt}{\rowfont\bfseries\color{uiText}#1:}\hspace{3pt}{\rowfont\color{uiTextLight}#2}\par}
\newcommand{\transrow}[5]{\par\noindent\badge[#1]{#2}\hspace{3.5pt}\iconarrow\hspace{3.5pt}\badge[#3]{#4}\hspace{4pt}{\detfont\itshape\color{uiTextMuted}(#5)}\par}
\newcommand{\donerow}[2]{\par\noindent\hangindent=11pt\hangafter=1\iconcheck{uiSuccess}\hspace{3.5pt}{\rowfont\bfseries\color{uiSuccess}#1 complete}\ifx\relax#2\relax\else\hspace{4pt}{\color{uiSep}$\cdot$}\hspace{4pt}{\rowfont\color{uiTextMuted}#2}\fi\par}
\newcommand{\progressbar}[2]{\par\vspace{2.5pt}\noindent{\detfont\color{uiTextMuted}Experiment #1 / #2}\par\vspace{1pt}\noindent\begin{tikzpicture}\fill[uiBorder,rounded corners=1.3pt] (0,0) rectangle (\linewidth,2.6pt);\fill[uiBlue,rounded corners=1.3pt] (0,0) rectangle ({#1/#2*\linewidth},2.6pt);\end{tikzpicture}\par}

\newcommand{\planaccent}{uiSuccess}
\newtcolorbox{plancard}[1]{enhanced,breakable,colback=uiBgSubtle,colframe=uiBorder,boxrule=0.4pt,arc=2.5pt,
  left=9pt,right=9pt,top=1pt,bottom=5pt,boxsep=0pt,before skip=3pt,after skip=3pt,
  title={#1},fonttitle=\hdrfont,coltitle=uiHeading,colbacktitle=uiBgSubtle,titlerule=0pt,
  toptitle=5pt,bottomtitle=0pt,lefttitle=9pt,fontupper=\chatlineskip\rowfont\color{uiTextLight}}
\newcommand{\plansec}[1]{\par\vspace{2.5pt}\noindent{\rowfont\bfseries\color{\planaccent}#1}\par\vspace{0.3pt}}
\newcommand{\planrow}[2]{\par\noindent\hangindent=5pt\hangafter=0\hspace*{5pt}{\bfseries\color{uiText}#1:} #2\par}
\newcommand{\planline}[1]{\par\noindent\hangindent=5pt\hangafter=0\hspace*{5pt}#1\par}
\newtcolorbox{logblock}[2]{enhanced,breakable,colback=uiBgSubtle,colframe=uiBorder,boxrule=0.4pt,arc=2.5pt,
  left=7pt,right=7pt,top=2pt,bottom=3pt,boxsep=0pt,before skip=2.5pt,after skip=2.5pt,
  title={\chevdown\hspace{4pt}{\detfont\bfseries\color{uiText}#1}\hfill{\detfont\color{uiTextMuted}#2}},
  colbacktitle=uiBgSubtle,titlerule=0pt,toptitle=2.5pt,bottomtitle=2.5pt,lefttitle=7pt,righttitle=7pt,
  fontupper=\chatlineskip\codefont\color{uiTextLight}}
\newcommand{\cl}[1]{\par\noindent\hangindent=8pt\hangafter=1 #1\par}

\newcommand{\actionbtn}[2]{\tcbox[on line,enhanced,colback=white,colframe=uiBorder,boxrule=0.5pt,arc=3.5pt,
  left=6pt,right=6pt,top=2pt,bottom=2pt,boxsep=0pt,fontupper=\btnfont\color{uiHeading}]{#1\hspace{4pt}#2}}
\newcommand{\actionbtnsel}[2]{\tcbox[on line,enhanced,colback=uiPrimary,colframe=uiPrimary,boxrule=0.5pt,arc=3.5pt,
  left=6pt,right=6pt,top=2pt,bottom=2pt,boxsep=0pt,fontupper=\btnfont\bfseries\color{white}]{#1\hspace{4pt}#2}}
\newcommand{\btnapprove}[1]{\ifnum#1=1 \actionbtnsel{\iconcheck{white}}{Approve}\else\actionbtn{\iconcheck{uiHeading}}{Approve}\fi}
\newcommand{\btnfeedback}[2]{\ifnum#1=1 \actionbtnsel{\iconmsg{white}}{#2}\else\actionbtn{\iconmsg{uiHeading}}{#2}\fi}
\newcommand{\btnreject}[1]{\ifnum#1=1 \actionbtnsel{\iconx{white}}{Reject (abort)}\else\actionbtn{\iconx{uiHeading}}{Reject (abort)}\fi}
\newcommand{\imgel}[2]{\par\vspace{2.5pt}\noindent\tcbox[on line,enhanced,colback=white,colframe=uiBorder,boxrule=0.4pt,arc=2pt,left=0pt,right=0pt,top=0pt,bottom=0pt,boxsep=0pt]{\includegraphics[width=#1]{\chatimgdir #2}}\par}
\newcommand{\elidefont}{\sffamily\fontsize{6.2}{6.2}\selectfont}
\newcommand{\elide}[1]{\hspace{0.25em}\tcbox[on line,enhanced,colback=uiBgSubtle,colframe=uiBgSubtle,
  boxrule=0pt,arc=2.5pt,borderline={0.4pt}{0pt}{uiTextMuted,dash pattern=on 2.5pt off 2pt},
  left=5pt,right=5pt,top=1.1pt,bottom=1.1pt,boxsep=0pt]{\elidefont\color{uiTextLight}#1}\hspace{0.25em}}
\newtcolorbox{omitted}{enhanced,colback=uiBgSubtle,colframe=uiBgSubtle,boxrule=0pt,arc=2.5pt,
  borderline={0.5pt}{0pt}{uiTextMuted,dash pattern=on 2.5pt off 2pt},
  left=8pt,right=8pt,top=5pt,bottom=5pt,boxsep=0pt,before skip=10pt,after skip=8pt,halign=center,
  fontupper=\chatlineskip\rowfont\color{uiTextLight}}

\usetikzlibrary{arrows.meta, positioning, shapes.geometric}
\newcommand{\partcheck}{\ooalign{\checkmark\cr\kern0.239em\raisebox{0.311em}{\rotatebox[origin=c]{-35}{\rule{0.50em}{0.050em}}}\hidewidth\cr}}

\journal{Elsevier}

\begin{document}

\begin{frontmatter}

\title{Gradient-aware AI agents for autonomous turbulence model discovery via differentiable CFD}

\author[cornellMAE]{Yi Liu}
\author[cornellMAE]{Xiantao Fan}
\author[cornellMAE]{Mingqi Xu}
\author[cornellMAE]{Jian-Xun Wang\corref{corr}}

\cortext[corr]{Corresponding author. Tel: +1 540 315 6512}
\ead{jw2837@cornell.edu}

\affiliation[cornellMAE]{organization={Sibley School of Mechanical and Aerospace Engineering, Cornell University},
            city={Ithaca},
            postcode={14853},
            state={NY},
            country={USA}}

\begin{abstract}
Turbulence modeling relies on human choices of closure form, physical inputs, and tunable parameters. Training closures \emph{a posteriori} is efficient only with gradients through long chaotic rollouts, which few solvers provide. Most \emph{a posteriori} approaches therefore resort to gradient-free optimization, and nearly all fix the closure form in advance, learning only its parameters. In this work, we introduce a gradient-aware agentic framework that poses closure discovery as a bilevel problem: a multi-agent large language model (LLM) system searches over closure forms, while a GPU-native differentiable computational fluid dynamics (CFD) solver supplies gradients, making differentiable \emph{a posteriori} learning an action of the autonomous AI agent. For each proposal, the agents construct a differentiable training program by embedding the analytical, neural, or hybrid closure in the solver, implementing a differentiable objective, and validating the gradient path. They then decide whether to evaluate the closure, compute its parameter sensitivities, or train it \emph{a posteriori}, select the optimization settings, and revise subsequent proposals using the simulation outcomes and returned gradients. Truncated-horizon gradients make this training usable under chaotic dynamics for closures with tens of thousands of parameters. Demonstrations include dissipation-law and subgrid-scale closure discovery in two-dimensional turbulence and wall-model discovery for large-eddy simulation of three-dimensional channel flow. With gradient-based training, the agent recovers reference coefficients that forward evaluation alone does not reach within the same budget, and autonomously discovers and trains analytical and neural closures that improve agreement with the targeted reference statistics. These results demonstrate the feasibility of end-to-end autonomous turbulence model development in time-resolved flows.
\end{abstract}

\end{frontmatter}

\section{Introduction}\label{introduction}

Turbulence governs the transport of momentum, heat, and mass in many engineering and natural flows. Resolving all turbulent scales remains prohibitively expensive at Reynolds numbers of practical interest, making turbulence models essential for predictive simulation. Such models are used, for example, in Reynolds-averaged Navier--Stokes (RANS) calculations, large-eddy simulation (LES), and simulations with unresolved near-wall layers. They represent quantities such as Reynolds and subgrid-scale (SGS) stresses and wall shear stress, as well as processes such as dissipation and backscatter. Widely used closures were developed through expert judgment: SGS models such as the Smagorinsky models \citep{smagorinsky1963,germano1991} and the Vreman model \citep{vreman2004}; RANS models such as $k$--$\varepsilon$ \citep{launder1974}, $k$--$\omega$ \citep{wilcox1988,menter1994}, and Spalart--Allmaras \citep{spalart1992}; and equilibrium and non-equilibrium wall models for LES \citep{piomelli2002,wang2002,yang2015,bose2018}. Their development involves selecting resolved inputs, invariants, functional forms, and coefficients, followed by calibration and testing against experimental or direct numerical simulation (DNS) data. Data-driven turbulence modeling has substantially expanded the available methods \citep{wang2017,duraisamy2019,brunton2020,vinuesa2022,sanderse2025}, but many design decisions remain with the modeler. Input features, model families, physical constraints, and network architectures are generally prescribed, while data determine the parameters within that structure. Automating the search over model forms under physical and numerical constraints remains an open challenge.

The training procedure is as important as the closure form. Most learned closures are trained \emph{a priori}, by regressing closure outputs onto targets extracted from reference fields \citep{beck2019,maulik2017,maulik2019,guan2022}. This decouples training from the solver, but a closure that accurately fits these targets may still be inaccurate or unstable in a simulation \citep{wu2026apost}. The training problem excludes the solver's numerical errors and the interactions between the closure, the resolved flow, and any other closures. \emph{A posteriori}, solver-in-the-loop training addresses this mismatch by evaluating the loss on the simulated flow, accounting for the closure's effects on the evolving flow and the statistics of interest \citep{um2020,list2022,shankar2025,frezat2022,kim2026,freitas2025,zhang2026diffwmles,fan2026differen}. Gradient-based training requires sensitivities of the flow solution to the closure parameters. These can be obtained using an adjoint solver \citep{sirignano2020,sirignano2023,kakka2026,dehtyriov2026} or a solver implemented in a differentiable programming framework such as JAX~\citep{jax2018github}. Solvers built using differentiable programming include JAX-CFD~\citep{kochkov2021}, Diff-FlowFSI~\citep{fan2026diff}, and DiFVM~\citep{du2026difvm}, among others \citep{bezgin2023,bezgin2025,holl2024,franz2026,saverio2026}. Most production CFD codes, however, are not differentiable, and adjoint solvers for unsteady, chaotic LES remain rare and costly to develop. Even when available, gradients through long chaotic rollouts can grow exponentially and require truncation or regularization to be useful \citep{wang2014,list2022}. Consequently, most \emph{a posteriori} closure learning relies on gradient-free optimization, which scales poorly with the number of parameters.
Whether gradient-based or gradient-free, these approaches fix the closure form in advance, and learning adjusts only its parameters within that prescribed form.

Other approaches automate parts of the closure-form search. Symbolic and sparse regression have produced interpretable RANS corrections, SGS models, and wall functions \citep{schmelzer2020,wu2025,cranmer2023,weatheritt2016,zhao2020,waschkowski2022,li2024,jakhar2026,ma2026rotation,fu2026symbolic}. Most are trained \emph{a priori} on reference data; CFD-driven variants evaluate candidates \emph{a posteriori}, but through gradient-free evolutionary search in steady RANS \citep{zhao2020,waschkowski2022,zemmour2026}. Reinforcement learning (RL) trains closures \emph{a posteriori} without solver gradients, typically by adapting a small number of coefficients of a classical SGS or wall model \citep{novati2021,bae2022,kurz2023,beck2023,zhou2024,vonbassewitz2025}. In both cases, the search space is fixed in advance through a term library, operator set, policy architecture, or action set.

Large language models (LLMs) offer another route. LLM-based agents can generate hypotheses, write and execute code, evaluate outcomes, and revise proposals, an approach termed autonomous scientific discovery \citep{wei2025agentic} or autonomous research \citep{ding2026survey} and implemented in systems often called AI scientists \citep{kitano2021,lu2026,mitchener2025,ding2026survey,feng2026primer}. Such systems have been demonstrated across mathematics and machine learning \citep{romeraparedes2024,novikov2025,lu2026,yin2026}, the natural sciences \citep{moss2025,cao2025,chen2026xray,boiko2023,bran2024,gottweis2026,ghareeb2026,swanson2025}, and engineering \citep{ma2024eureka,ni2024mechagents,jain2026autoresearch}. In fluid mechanics, most agentic systems to date automate forward simulation, translating natural-language descriptions into CFD cases and handling execution, error repair, and post-processing \citep{chen2024,chen2025,pandey2025,feng2026openfoam,yue2026,fan2026,xu2025,yang2026,dong2026,fan2026phynikce,lee2026review}; others address turbomachinery design \citep{du2026}, laboratory experimentation \citep{feng2025}, and flow control \citep{sun2026}.

For turbulence model discovery, LLMs have been used to propose closure forms. AutoTurb \citep{zhang2025} guides LLM proposals of algebraic closures by evolutionary search and evaluates them in the solver; PiT-PO \citep{wang2026} fine-tunes the LLM by policy optimization with rewards from fitting reference data; and AI CFD Scientist \citep{somasekharan2026}, a general CFD research agent built on OpenFOAM, patches model libraries and discovered a Spalart–Allmaras correction for periodic-hill flow. PhysMiner \citep{chen2026physminer} derives an SGS model from LLM reasoning over velocity-gradient statistics, and \citet{yang2025} obtained LLM-proposed wall-model modifications for pressure-gradient, rotating, and rough-wall flows, implemented and evaluated \emph{a posteriori} by human researchers. General equation-discovery systems such as LLM-SR \citep{shojaee2025} and LLM4ED \citep{du2024} similarly combine program-based proposals with parameter fitting. Because candidate forms are expressed as code informed by the modeling literature, these systems can search beyond a fixed coefficient set and, in some cases, beyond a predefined term library \citep{lu2026,somasekharan2026}. Their solver feedback, however, is limited to evaluation results such as scalar metrics, logs, or field snapshots, and parameter fitting relies on static reference data or repeated forward simulations rather than gradients through the flow. Tuning even a few coefficients can then require many costly simulations, and high-dimensional neural or hybrid closures cannot practically be trained \emph{a posteriori} within the loop. Moreover, autonomous searches have mostly targeted steady RANS, while three-dimensional LES evaluations of LLM-proposed closures have required separate human implementation and testing \citep{yang2025,chen2026physminer}. Whether the complete discovery cycle, from proposal through implementation and training to revision, can operate autonomously in time-resolved three-dimensional turbulence therefore remains open.

Differentiable solver-in-the-loop learning and agentic model-form search address complementary aspects of closure discovery: the former trains closures within a prescribed form using gradients obtained from differentiable solvers, while the latter explores model forms using evaluation feedback. Related combinations have appeared outside fluid mechanics: the Scientific Generative Agent \citep{ma2024} couples LLM proposals with differentiable simulation for constitutive-law discovery and molecular design, and \citet{tacke2026} combine LLM-driven constitutive-model design with parameter fitting and constraint validation. Turbulence closure discovery poses challenges absent from these settings: gradients must be propagated through long rollouts of chaotic flow with noisy objectives, closures may carry tens of thousands of parameters, and each proposal must be translated into a solver-compatible differentiable program before it can be trained.

This work introduces a gradient-aware agentic framework for turbulence model discovery, in which differentiable \emph{a posteriori} learning becomes an action of an autonomous agent. An LLM-based multi-agent system searches over closure forms and physical inputs, while a GPU-native differentiable CFD solver supplies gradients through time-resolved turbulence simulations. The framework is gradient-aware in three respects: the agents construct and validate the differentiable training program for each proposal, decide when and how to invoke gradient-based training, and use the returned sensitivities to revise subsequent proposals. This pairs discrete exploration of closure forms with continuous parameter training, making analytical, neural, and hybrid closures accessible within one workflow. Gradient-based training reduces reliance on trial-and-error parameter adjustment, while automated implementation, validation, and failure diagnosis remove the manual work between modeling ideas and numerical evaluation. To our knowledge, this is the first agentic AI framework to close the full discovery cycle, from proposal through differentiable \emph{a posteriori} training to revision, autonomously in time-resolved turbulence. We demonstrate the framework on three canonical problems with well-characterized physics and reference solutions, which allow both the operation of the discovery loop and the physical behavior of the resulting closures to be assessed. Table~\ref{tab:positioning} compares its capabilities with those of the approaches reviewed above.
\begin{table}[!t]
\centering
\footnotesize
\caption{Comparison of the present framework with existing closure-discovery approaches in turbulence modeling. Columns denote method families or individual systems. \checkmark: present; \protect\partcheck: partial (see notes); --: absent or not applicable.}
\label{tab:positioning}
\setlength{\tabcolsep}{5pt}
\renewcommand{\arraystretch}{1.0}
\begin{tabular}{@{}lccccccccc@{}}
\toprule
& \rotatebox{90}{\makecell[l]{Symbolic regression\\\citep{schmelzer2020,wu2025,weatheritt2016,zhao2020,waschkowski2022,li2024,jakhar2026,ma2026rotation,fu2026symbolic}}}
& \rotatebox{90}{\makecell[l]{RL closure\\discovery \citep{novati2021,bae2022,kurz2023,beck2023,zhou2024,vonbassewitz2025}}}
& \rotatebox{90}{\makecell[l]{Differentiable/adjoint\\learning \citep{um2020,list2022,shankar2025,frezat2022,kim2026,freitas2025,fan2026differen,zhang2026diffwmles,sirignano2020,sirignano2023,kakka2026,dehtyriov2026,saverio2026}}}
& \rotatebox{90}{\makecell[l]{Yang et al. \citep{yang2025}}}
& \rotatebox{90}{\makecell[l]{AutoTurb \citep{zhang2025}}}
& \rotatebox{90}{\makecell[l]{PiT-PO \citep{wang2026}}}
& \rotatebox{90}{\makecell[l]{AI CFD Scientist \citep{somasekharan2026}}}
& \rotatebox{90}{\makecell[l]{PhysMiner \citep{chen2026physminer}}}
& \rotatebox{90}{\makecell[l]{\textbf{This work}}} \\
\midrule
Open-space closure proposal$^{a}$ & \partcheck$^{b}$ & -- & -- & \checkmark & \partcheck$^{b}$ & \partcheck$^{b}$ & \checkmark & \checkmark & \pmb{\checkmark} \\
\addlinespace[2.5pt]
Neural/hybrid closures & -- & \checkmark$^{c}$ & \checkmark & \partcheck$^{d}$ & -- & -- & -- & -- & \pmb{\checkmark} \\
\addlinespace[2.5pt]
\emph{A posteriori} parameter training & \partcheck$^{e}$ & \checkmark & \checkmark & -- & \partcheck$^{e}$ & --$^{f}$ & \partcheck$^{e}$ & -- & \pmb{\checkmark} \\
\addlinespace[2.5pt]
Evaluation in 3D turbulence$^{g}$ & \checkmark & \checkmark & \checkmark & \checkmark & -- & -- & -- & \checkmark & \pmb{\checkmark} \\
\addlinespace[2.5pt]
Gradient feedback to proposer$^{h}$ & -- & -- & -- & -- & -- & -- & -- & -- & \pmb{\checkmark} \\
\addlinespace[2.5pt]
Automated code validation & -- & -- & -- & -- & \partcheck$^{i}$ & -- & \checkmark & -- & \pmb{\checkmark} \\
\addlinespace[2.5pt]
Automated propose--evaluate loop & \checkmark$^{j}$ & -- & -- & -- & \checkmark & \checkmark & \checkmark & -- & \pmb{\checkmark} \\
\bottomrule
\end{tabular}
\par\smallskip
\begin{minipage}{\textwidth}\scriptsize
$^{a}$Free-form expressions or code; neural forms are assessed separately.
$^{b}$Search restricted by prescribed libraries, operator sets, or tensor templates.
$^{c}$Neural policies, mostly supplying a coefficient to a classical closure.
$^{d}$One hybrid model, LLM-advised, trained by the authors on static data.
$^{e}$Coefficients adjusted by evolution or re-proposal, without a separate parameter optimizer.
$^{f}$Fitted \emph{a priori}; solver runs, if any, only rank candidates.
$^{g}$Time-resolved simulation of 3D turbulence, as opposed to steady RANS or 2D flow.
$^{h}$Simulation gradients returned to the model-form proposer.
$^{i}$Feasibility screen only.
$^{j}$Automated model-form search and numerical evaluation \citep{zhao2020,waschkowski2022}; final selection may be manual.
\end{minipage}
\end{table}
Making differentiable \emph{a posteriori} learning available to an agent requires translating each proposal into a differentiable training program. Closure parameters must be carried in the solver state, every operation along the computational path must support reverse-mode differentiation, and the differentiated horizon must be short enough to yield useful gradients under chaotic dynamics. The framework automates this translation for analytical, neural, and hybrid closures through a shared differentiable interface, and validates the gradient path by static checks and a short trial run before full execution. When the objective is not specified, it is derived from the research question and implemented as a differentiable loss, which is checked against a direct evaluation of the target statistic before use. For each experiment, the agents then choose among forward evaluation, gradient sensitivity, and full \emph{a posteriori} optimization, balancing the information gained against computational cost: forward evaluation screens candidates, sensitivities identify influential parameters and indicate descent directions, and optimization trains high-dimensional closures with the solver in the loop.
The main contributions are as follows.
\begin{itemize}
\item A gradient-aware agentic framework that poses turbulence model discovery as a bilevel search, in which LLM agents explore closure forms and physical inputs while a differentiable CFD solver supplies gradients for \emph{a posteriori} training. Differentiable \emph{a posteriori} learning thereby becomes an action of the agent, extending closure development beyond a prescribed model form and reducing reliance on trial-and-error tuning through numerous forward simulations.
\item Automated construction and validation of differentiable training programs for analytical, neural, and hybrid closures, including high-\hspace{0pt}dimensional ones, trained under chaotic dynamics.
\item A dual use of gradients from differentiable programming across both levels of the search. Within an experiment, they drive \emph{a posteriori} optimization of the closure parameters; across experiments, the returned sensitivities and training histories inform the proposer’s choice of solver mode, optimization settings, and subsequent closure revisions.
\end{itemize}

The remainder of the paper is organized as follows. Section~\ref{problem-formulation} defines the closure-discovery problem, the \emph{a posteriori} objective, rollout gradients, and the three solver modes. Section~\ref{gradient-aware-agentic-framework} describes the agentic framework. Section~\ref{numerical-experiments} presents dissipation-law and SGS closure discovery in two-dimensional turbulence, wall-model discovery for wall-modeled LES (WMLES) of turbulent channel flow, an analysis of proposals across runs, and an assessment of the role of gradients through a solver-mode comparison and a gradient-feedback ablation. Section~\ref{conclusions-and-outlook} presents conclusions and directions for further work.

\section{Problem formulation}\label{problem-formulation}

\subsection{Governing equations and closure terms}\label{closure-terms}

We consider incompressible turbulent flows in which limited grid resolution or omitted physical processes require a closure model. The resolved velocity \(\bm{u}(\bm{x},t)\) satisfies the incompressible Navier--Stokes equations augmented by a closure term,
\begin{equation}
\partial_t \bm{u} + (\bm{u}\cdot\nabla)\bm{u} = -\frac{1}{\rho}\nabla p + \nu\nabla^2\bm{u} + \bm{f} + \mathcal{M}(\bm{u};\bm{\theta}),
\qquad \nabla\cdot\bm{u} = 0,
\label{eq:resolved-ns}
\end{equation}
where \(p\) is the pressure, \(\rho\) is the density, \(\nu\) is the kinematic viscosity, \(\bm{f}\) is a body force per unit mass, and \(\mathcal{M}\) models the effect of unresolved dynamics on the resolved field. The closure depends on the resolved fields and a parameter vector \(\bm{\theta}\in\mathbb{R}^{n_\theta}\). We use \(\bm{u}\) as its argument for brevity, although it may depend on any resolved field, including pressure, and this dependence may be local, through pointwise values and their derivatives, or nonlocal, through the full field. We distinguish the closure's \emph{form} \(\mathcal{M}\), which specifies its functional dependence on the resolved fields, from its \emph{parameters} \(\bm{\theta}\), which may comprise a few scalar coefficients or the weights of a neural network.

A closure may also represent unresolved physics through a boundary condition, such as a wall-stress condition, synthetic-turbulence inflow, or roughness model. It then prescribes a boundary value or flux on a portion \(\Gamma\) of the boundary,
\begin{equation}
\mathcal{B}(\bm{u},p) = \mathcal{M}_{\Gamma}(\bm{u};\bm{\theta}) \quad \text{on } \Gamma,
\label{eq:closure-bc}
\end{equation}
where \(\mathcal{B}\) extracts the prescribed boundary quantity, either a value or a flux of the resolved fields. For both types of closures, the problem specifies where the closure acts, its admissible inputs, its output, and any physical constraints on that output. The closure form and parameters are to be discovered within these specifications.

\subsection{A posteriori objective and the discovery problem}\label{discovery-problem}

We assess a closure \emph{a posteriori} using statistics of the flow obtained by advancing Eq.~\eqref{eq:resolved-ns} in time. In contrast, \emph{a priori} assessment compares the closure output with the unclosed term evaluated on reference fields. The closed equations are advanced until the flow is statistically stationary, then over an evaluation horizon divided into \(M\) equal windows. Let \(\mathcal{S}\) denote a resolved-flow statistic accumulated over all \(M\) windows, and \(\mathcal{S}_{\mathrm{ref}}\) the corresponding statistic from a reference, such as a higher-fidelity simulation or experimental data. The objective is
\begin{equation}
\mathcal{J}(\mathcal{M},\bm{\theta}) = d\big(\mathcal{S}(\mathcal{M},\bm{\theta}),\,\mathcal{S}_{\mathrm{ref}}\big),
\label{eq:objective}
\end{equation}
where \(d\) measures the discrepancy between statistics. The statistic and discrepancy measure are specified as part of the problem or, when unspecified, derived by the agent from the stated goal (Section~\ref{differentiable-objective-from-the-research-question}). Because \(\mathcal{J}\) is evaluated on the simulated flow, it reflects the combined effects of the closure, numerical scheme, and resolution. Because \(\mathcal{S}\) is accumulated over a finite horizon of a chaotic flow, \(\mathcal{J}\) is also a random variable that depends on the flow realization. We refer to its variability across independent realizations under a fixed closure as the noise level of the objective; differences between closures below this level are not considered significant.

Closure discovery is then a bilevel optimization problem
\begin{equation}
\min_{\mathcal{M}\in\mathcal{H}}\;\min_{\bm{\theta}\in\Theta_{\mathcal{M}}}\;\mathcal{J}(\mathcal{M},\bm{\theta}),
\label{eq:two-level}
\end{equation}
where \(\mathcal{H}\) is the space of admissible closure forms and \(\Theta_{\mathcal{M}}\) is the parameter space of a given form. Admissible forms must be implementable in the solver, respect the settings designated as fixed in the problem specification, and satisfy the constraints in Section~\ref{closure-terms}. The outer search is discrete: its candidates are functional forms, and changing the form may change both the dimension and interpretation of \(\bm{\theta}\). The inner optimization is continuous, with gradients provided by the differentiable solver. The framework in Section~\ref{gradient-aware-agentic-framework} assigns model-form proposals to an LLM, informed by its learned modeling knowledge and previous simulation outcomes, and parameter optimization and sensitivity evaluation to the solver. Neither level is solved to optimality: the outer search is a heuristic, history-conditioned exploration of \(\mathcal{H}\) by the LLM, and the inner problem is approximated by a finite number of gradient steps on a noisy objective.

\subsection{Differentiable solver and rollout gradients}\label{rollout-gradients}

The forward model is a flow solver implemented in a differentiable programming framework, allowing its operations, including the closure, to be differentiated by reverse-mode automatic differentiation. Writing one time step as \(\bm{u}_{n+1} = \Phi(\bm{u}_n;\bm{\theta})\), the composition of the \(N\) steps in the evaluation horizon is also differentiable. For a fixed initial state and an objective whose parameter dependence enters through the flow states, the chain rule gives:
\begin{equation}
\frac{\mathrm{d}\mathcal{J}}{\mathrm{d}\bm{\theta}}
= \sum_{n=1}^{N} \frac{\partial\mathcal{J}}{\partial\bm{u}_n}\,\frac{\mathrm{d}\bm{u}_n}{\mathrm{d}\bm{\theta}},
\qquad
\frac{\mathrm{d}\bm{u}_{n}}{\mathrm{d}\bm{\theta}}
= \frac{\partial\Phi}{\partial\bm{u}}\bigg|_{n-1}\frac{\mathrm{d}\bm{u}_{n-1}}{\mathrm{d}\bm{\theta}}
+ \frac{\partial\Phi}{\partial\bm{\theta}}\bigg|_{n-1}.
\label{eq:rollout-gradient}
\end{equation}
Intermediate states are stored at selected time steps and recomputed during the reverse pass to limit memory use. Closure parameters enter the derivative through \(\partial\Phi/\partial\bm{\theta}\), so the same mechanism applies to scalar coefficients, analytical expressions, and neural networks.

Turbulent flow is chaotic: state perturbations can grow exponentially in time, as can the sensitivities \(\mathrm{d}\bm{u}_n/\mathrm{d}\bm{\theta}\) in Eq.~\eqref{eq:rollout-gradient}. Consequently, the gradient of a finite-time average need not converge to the derivative of the corresponding long-time statistic as the averaging horizon increases \citep{wang2014,blonigan2018}. Shortening the differentiated horizon limits this growth at the cost of bias, a common approach in solver-in-the-loop closure learning \citep{list2022,hickling2026,fan2026differen}. Training and evaluation use the same objective over all \(M\) windows, but derivatives are propagated only through the last \(K\) windows. The earlier states, including their contributions to the accumulated statistics and the initial state of the differentiated rollout, are held fixed during differentiation:
\begin{equation}
\bm{g} = \frac{\mathrm{d}\mathcal{J}}{\mathrm{d}\bm{\theta}}\bigg|_{\bm{u}_n\ \mathrm{fixed},\;0\le n\le N_{M-K}},
\label{eq:truncated-gradient}
\end{equation}
where \(N_{M-K}\) is the index of the last time step before the differentiated windows; for \(K=M\), \(N_0=0\) and only the initial state is held fixed. Truncation changes the gradient but leaves the objective value unchanged. Before evaluation, a forward-only warm-up brings the flow to a statistically stationary state under the current closure. Gradients are therefore evaluated after the initial transient. The warm-up duration, number of evaluation windows \(M\), and number of differentiated windows \(K\) are specified for the study or chosen by the agent when unspecified (Section~\ref{differentiable-objective-from-the-research-question}). These settings remain fixed across experiments in a run to ensure comparable objective values.

\subsection{Solver modes}\label{feedback-modes}

The agent interacts with the solver through a sequence of experiments, each comprising one proposal, its implementation, and its evaluation. Experiment \(i\) begins with a proposal \((\mathcal{M}_i,\bm{\theta}_i,m_i,\bm{\xi}_i) = \mathcal{P}(\mathcal{R}_{<i})\) from the LLM proposer \(\mathcal{P}\), where \(m_i\) is the solver mode, \(\bm{\xi}_i\) collects the optimization settings when training is requested, and \(\mathcal{R}_{<i}\) records all previous proposals and solver feedback. The three solver modes constitute the actions through which the agent uses gradients; they differ in the computation performed within each experiment and the feedback \(\mathcal{F}_i\) returned to the proposer.
\emph{Forward evaluation} computes the objective, \emph{gradient sensitivity} additionally computes parameter gradients, and \emph{a posteriori optimization} updates parameters through repeated objective and gradient evaluations. All modes return diagnostics \(\mathcal{D}_i\): the run status (completed, diverged, or failed), statistics used in the objective, and summary quantities for the closure. Forward evaluation returns the objective and these diagnostics,
\begin{equation}
\mathcal{F}_i = \big\{\mathcal{J}(\mathcal{M}_i,\bm{\theta}_i),\ \mathcal{D}_i\big\},
\label{eq:mode0}
\end{equation}
corresponding to the black-box feedback used in existing agentic CFD systems. In forward evaluation and gradient sensitivity, parameters remain fixed within an experiment; changes between experiments are made by the proposer. Gradient sensitivity computes the truncated rollout gradient in Eq.~\eqref{eq:truncated-gradient} and returns
\begin{equation}
\mathcal{F}_i = \big\{\mathcal{J}(\mathcal{M}_i,\bm{\theta}_i),\ \mathcal{G}\big(\bm{g}(\mathcal{M}_i,\bm{\theta}_i)\big),\ \mathcal{D}_i\big\},
\label{eq:mode1}
\end{equation}
where \(\mathcal{G}\) reports signed derivatives for scalar parameters and gradient norms for parameter arrays, including neural-network weights. The proposer can use this information to adjust parameter values or revise the closure form.

In \emph{a posteriori} optimization, the solver performs up to \(E\) gradient-based optimization epochs, with the epoch budget \(E\), learning rate \(\eta\), and optimizer specified by \(\bm{\xi}_i\). Let \(E_i\le E\) denote the number of epochs completed in experiment \(i\). For gradient descent, the update takes the form
\begin{equation}
\bm{\theta}^{(e+1)} = \bm{\theta}^{(e)} - \eta\,\bm{g}^{(e)},\qquad e = 0,\ldots,E_i-1,
\label{eq:mode2-update}
\end{equation}
starting from the proposed parameters, \(\bm{\theta}^{(0)} = \bm{\theta}_i\), or previously trained parameters if requested. Each epoch evaluates the objective \(\mathcal{J}^{(e)}\) and truncated gradient \(\bm{g}^{(e)}\) at \(\bm{\theta}^{(e)}\) before updating the parameters. The lowest-loss epoch \(e_i^{\star}\) determines the retained parameter vector \(\bm{\theta}_i^{\star}\). Feedback includes this objective value, the training loss history, retained parameters, gradient information from the best epoch, and diagnostics:
\begin{equation}
\begin{aligned}
e_i^{\star} &= \operatorname*{arg\,min}_{0\le e<E_i}\mathcal{J}^{(e)},
\qquad \bm{\theta}_i^{\star} = \bm{\theta}^{(e_i^{\star})},\\
\mathcal{F}_i &= \big\{\mathcal{J}(\mathcal{M}_i,\bm{\theta}_i^{\star}),\ \mathcal{J}^{(0:E_i-1)},\ \bm{\theta}_i^{\star},\ \mathcal{G}\big(\bm{g}^{(e_i^{\star})}\big),\ \mathcal{D}_i\big\}.
\end{aligned}
\label{eq:mode2}
\end{equation}
A checkpoint of \(\bm{\theta}_i^{\star}\) is also saved for use in later experiments. These modes differ in computational cost as well as the feedback returned. Gradient sensitivity adds one reverse pass to a forward evaluation, while \emph{a posteriori} optimization repeats forward and reverse passes over \(E_i\) epochs. Forward evaluation relies on the proposer's ability to tune parameters from objective values; \emph{a posteriori} optimization enables efficient training of high-dimensional closures such as neural networks. In this mode, gradients act at both levels of Eq.~\eqref{eq:two-level}: they update \(\bm{\theta}\) within the experiment, and, through \(\mathcal{G}\) and the loss history, inform the proposer's subsequent choices of \(\mathcal{M}\), \(m\), and \(\bm{\xi}\). The agent's mode selection is described in Section~\ref{gradient-aware-feedback-modes}, and the contribution of each mode is assessed in Section~\ref{ablation-of-the-feedback-modes}.

\section{Gradient-aware agentic framework}\label{gradient-aware-agentic-framework}

\subsection{System overview}\label{system-overview}

\begin{figure}[!htp]
  \centering
  \includegraphics[width=\textwidth]{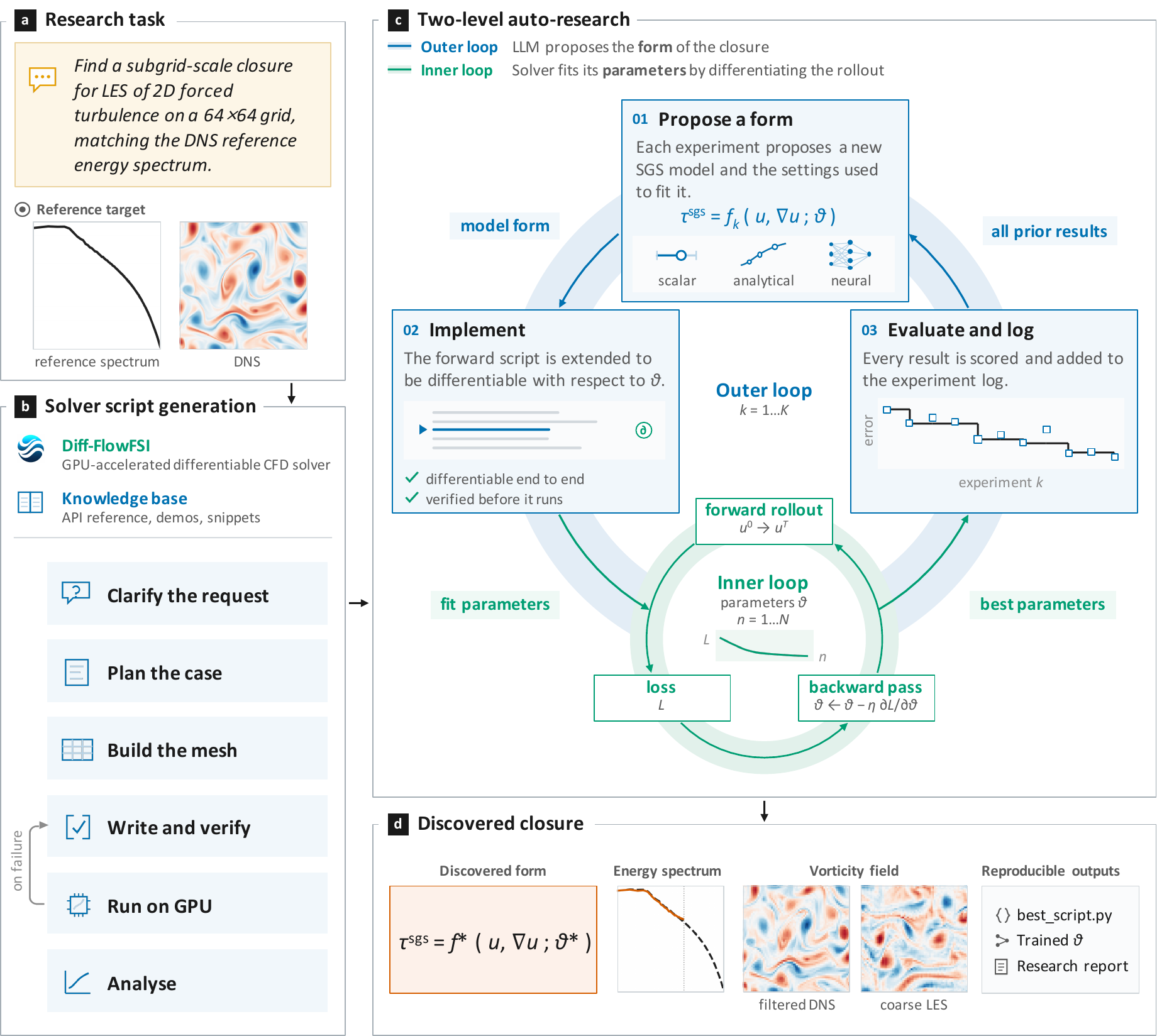}
  \caption{Overview of the gradient-aware agentic framework, illustrated on a subgrid-scale model discovery task.}
  \label{fig:framework}
\end{figure}

The framework has three main components, as illustrated in Fig.~\ref{fig:framework}: a differentiable flow solver, a knowledge base documenting its use, and a multi-agent system that conducts the research. The solver, Diff-FlowFSI \citep{fan2026diff}, is a GPU-native finite-volume CFD solver implemented in JAX. It supports staggered and collocated grids, a fractional-step projection method with a pressure Poisson solve, explicit or semi-implicit time integration, and immersed boundary and fluid--structure interaction capabilities. Compatible closures can be differentiated through the rollout (Section~\ref{rollout-gradients}), with gradients available both to the proposer and to the optimizer. The knowledge base contains the solver's API reference, demonstration scripts, and code snippets, which the LLM consults when writing solver code. The multi-agent system is organized as a state graph: each node represents an LLM call with a specific role, a deterministic program step, or an optional human review point. Conditional edges implement retries and repairs (\ref{app:workflow-graph}). The same LLM serves every role, with agents distinguished by role-specific prompts, tools, and inputs; the design is independent of the LLM provider.

The graph has two parts. The simulation pipeline (Fig.~\ref{fig:framework}b) converts a natural-language flow description into a working solver script. It clarifies the request, plans the case, builds the mesh, adapts a suitable demonstration script, checks and runs it on a GPU, and diagnoses and repairs failures. Users may also supply an existing simulation script. The auto-research loop (Fig.~\ref{fig:framework}c) implements the bilevel search of Section~\ref{discovery-problem} with the solver modes of Section~\ref{feedback-modes}. It starts from the base simulation and a research question specifying the target statistic, reference, and permitted modifications (Fig.~\ref{fig:framework}a). Its outputs are a research report, a self-contained script implementing the best closure, and trained parameters (Fig.~\ref{fig:framework}d). An initializer configures the study and evaluates the base simulation. Each experiment then passes through five roles: a proposer selects a closure form, parameters, solver mode, and optimization settings; a modifier implements the proposal; a validator checks the script statically and in a short trial run; an executor runs it on the GPU; and an analyzer collects the objective, gradients, training results, and diagnostics in the experiment record. A synthesizer writes the report when the experiment budget is exhausted. The proposer conducts the outer search in Eq.~\eqref{eq:two-level}, while the differentiable solver performs the inner parameter optimization whenever \emph{a posteriori} optimization is requested.

Human oversight is optional. The simulation pipeline can pause for review of the plan, mesh, and generated script, and again after the baseline run. At each point, the user may approve, revise, or reject the result. With these checks disabled, as in Section~\ref{numerical-experiments}, the framework proceeds autonomously. A run with review points enabled is reproduced in \ref{app:chat-transcript}. Experiment artifacts, including scripts, solver outputs, gradients, training histories, checkpoints, and structured summaries, are saved to disk to support resumption and reproducibility. The framework is implemented in Python using LangGraph \citep{langgraph}; closures are Equinox modules \citep{kidger2021equinox} trained with Optax \citep{deepmind2020jax}. It is also available as a Model Context Protocol (MCP) server for use by other agentic systems. All results reported here were obtained with the framework operating independently.

\subsection{From research question to a differentiable objective}
\label{differentiable-objective-from-the-research-question}

Before the first experiment, the agent reads the research question and base simulation to identify the permitted modifications, fixed settings, and evaluation criteria. It also determines how the closure enters the solver, for example through a body force, an interior eddy viscosity, or a wall-stress condition. The warm-up, averaging, and differentiated durations in Section~\ref{rollout-gradients} are taken from the research question or, when unspecified, selected using the characteristic flow time scales. They remain fixed throughout the study.

The agent constructs the objective by writing a program that computes the target statistic, compares it with the reference, and returns \(\mathcal{J}\). It then translates this program into a differentiable function of the simulated trajectory. On the baseline simulation, it verifies that both implementations return the same value and that the gradient is finite. The same objective function and window aggregation are used for training and for ranking experiments.

\subsection{Closure proposal, implementation, and validation}\label{closure-proposal-and-implementation}

The proposer reads the research question, current base simulation, and record \(\mathcal{R}_{<i}\) of previous experiments, including the code of the best closures. It then specifies the next experiment: the closure form in physical terms, resolved inputs, trainable parameters, initial parameter values, and solver mode $m_i$. For training, it also specifies $\bm{\xi}_i$: the epoch count \(E\), learning rate \(\eta\), and optimizer settings. The proposer is instructed to explore all admissible closure families rather than repeatedly refine a single family.

The modifier implements the proposal in a solver script, consulting the solver's functions, source code, and reference implementations. All closure families use the same interface: trainable parameters are collected in a module carried with the flow state, and the closure reads this module and the resolved fields at each solver call. The loss function inserts the module, advances the solver over the differentiated horizon, and returns the objective. The LLM writes the closure, while templates implement the warm-up, truncated-horizon rollout, gradient evaluation or optimization loop, and output routines. These templates use the study settings specified by the user or chosen by the agent, together with the solver mode selected by the proposer for each experiment. The agent is thus responsible for the closure, its coupling to the solver, and the differentiable objective, while the templates ensure that every closure is trained and evaluated by the same procedure, so that objective values remain comparable across experiments.

Before full execution, the script undergoes static checks of its code, solver calls, fixed settings, and compatibility with automatic differentiation. A short trial run then checks for crashes and non-finite or vanishing gradients. Failed checks return the script to the modifier for repair. A validated experiment is run in full, and its outcome, including any crash or divergence, is recorded with the analyzer's diagnosis for the proposer. Failed experiments thus provide evidence about the numerical behavior of a closure family. Whenever an experiment improves the best objective, its closure is incorporated into the base simulation for subsequent experiments.

\subsection{Solver-mode selection and training under chaotic dynamics}\label{gradient-aware-feedback-modes}

The three solver modes in Section~\ref{feedback-modes} differ in the computation performed after closure integration: forward evaluation, a single evaluation of the objective and gradient, or the optimization loop in Eq.~\eqref{eq:mode2-update}. The proposer selects the mode for each experiment. It is guided to balance the expected benefit of each mode against its computational cost, using less expensive evaluations to inform exploration before investing in parameter optimization. Neural closures are always trained, since a gradient at random initialization provides limited information about their attainable performance. Users may restrict the available modes, as in the mode comparison in Section~\ref{ablation-of-the-feedback-modes}.

Training under chaotic dynamics requires additional safeguards. Each epoch evaluates a different realization of the chaotic flow, producing a noisy loss. The parameters from the lowest-loss epoch are retained, and gradients are clipped to limit the effect of occasional large values.
When available, objective variations over later warm-up windows provide a heuristic estimate of the noise level defined in Section~\ref{discovery-problem}, assuming transient effects have sufficiently diminished. This informs early stopping and the proposer's assessment of differences between experiments.

\section{Numerical experiments}\label{numerical-experiments}

We demonstrate the framework on three closure-discovery problems, each using a different interface in the resolved equations: a dissipation law acting on the velocity in forced two-dimensional turbulence (Case~1), an SGS stress for the same flow on a coarse grid (Case~2), and a wall-stress boundary condition for three-dimensional WMLES of turbulent channel flow (Case~3). These canonical problems allow us to assess the complete autonomous discovery cycle in controlled settings and evaluate the resulting closures against established physical mechanisms and reference solutions. Although the agent can generate the base simulation, all runs within each case use the same base script for comparability. The agent receives this script and a research question specifying the target statistic and relevant constraints, then runs autonomously under a fixed experiment budget. GPT-5-mini is used for all agent roles. Table~\ref{tab:cases} summarizes the configurations and evaluation settings. All three cases involve time-resolved chaotic turbulence evaluated within the autonomous loop. Each uses the objective in Eq.~\eqref{eq:objective}, with a prescribed statistic and discrepancy measure evaluated over time windows after the flow reaches statistical stationarity, as described in Section~\ref{rollout-gradients}. We also examine flow statistics not included in the training objectives, providing a broader assessment of the physical behavior of the resulting closures. Section~\ref{ablation-of-the-feedback-modes} compares the solver modes and separately tests the effect of gradient feedback to the proposer.

\begin{table}[!t]
\centering
\footnotesize
\caption{Configuration of the three discovery studies.}
\label{tab:cases}
\scriptsize
\setlength{\tabcolsep}{4pt}
\renewcommand{\arraystretch}{1.05}
\begin{tabular}{@{}>{\raggedright\arraybackslash}p{0.175\textwidth}>{\raggedright\arraybackslash}p{0.255\textwidth}>{\raggedright\arraybackslash}p{0.255\textwidth}>{\raggedright\arraybackslash}p{0.255\textwidth}@{}}
\toprule
& Case 1: dissipation law & Case 2: SGS closure & Case 3: wall model \\
\midrule
\multicolumn{4}{@{}l}{\emph{Problem}} \\
Base model & Forced 2D turbulence on a \(256^2\) grid; linear drag \(\alpha = 0.075\) & Forced 2D turbulence on a \(64^2\) grid; linear drag \(\alpha = 0.15\), no SGS closure & Channel WMLES on a \(48\times24\times48\) grid; equilibrium wall model, Vreman SGS \\
\addlinespace[3pt]
Reference & Same flow with drag \(\alpha = 0.15\) and hyperviscosity & Same physics resolved on a \(512^2\) grid & DNS at \(Re_\tau = 392\) \citep{moser1999} \\
\addlinespace[3pt]
Objective & Log-spectral error & Log-spectral error & Mean-velocity error in wall units \\
\addlinespace[3pt]
Closure interface & Body force on the resolved field & Eddy viscosity & Wall-stress magnitude \\
\addlinespace[3pt]
Agent may modify & Dissipation form, inputs, and parameters; forcing parameters & Form, inputs, and parameters of \(\nu_t\) & Form, inputs, and parameters of \(\tau_w\) \\
\addlinespace[3pt]
Held fixed & Grid, domain, viscosity, numerics & Grid, domain, viscosity, forcing, drag, numerics & Grid, domain, viscosity, bulk flow rate, SGS model, numerics \\
\addlinespace[3pt]
Closure-family regimes & Analytical-only, neural-only, unrestricted & Analytical-only, neural-only, unrestricted & Unrestricted \\
\addlinespace
\multicolumn{4}{@{}l}{\emph{Evaluation}} \\
Number of runs & 3 & 3 & 1 \\
\addlinespace[3pt]
Experiments per run & 30 & 30 & 30 \\
\addlinespace[3pt]
Window length (time units) & 1 & 1 & 16 \\
\addlinespace[3pt]
Windows averaged, \(M\) & 5 & 5 & 5 \\
\addlinespace[3pt]
Windows differentiated, \(K\) & 5 & 5 & 1 \\
\addlinespace[3pt]
Time per run & 8--10 h & 4--11 h & 82 h \\
\bottomrule
\end{tabular}
\end{table}

\subsection{Case 1: dissipation-law discovery in 2D turbulence}\label{case-1-dissipation-law-discovery-in-2d-hit}
\begin{figure}[!tp]
  \centering
  \includegraphics[width=\textwidth]{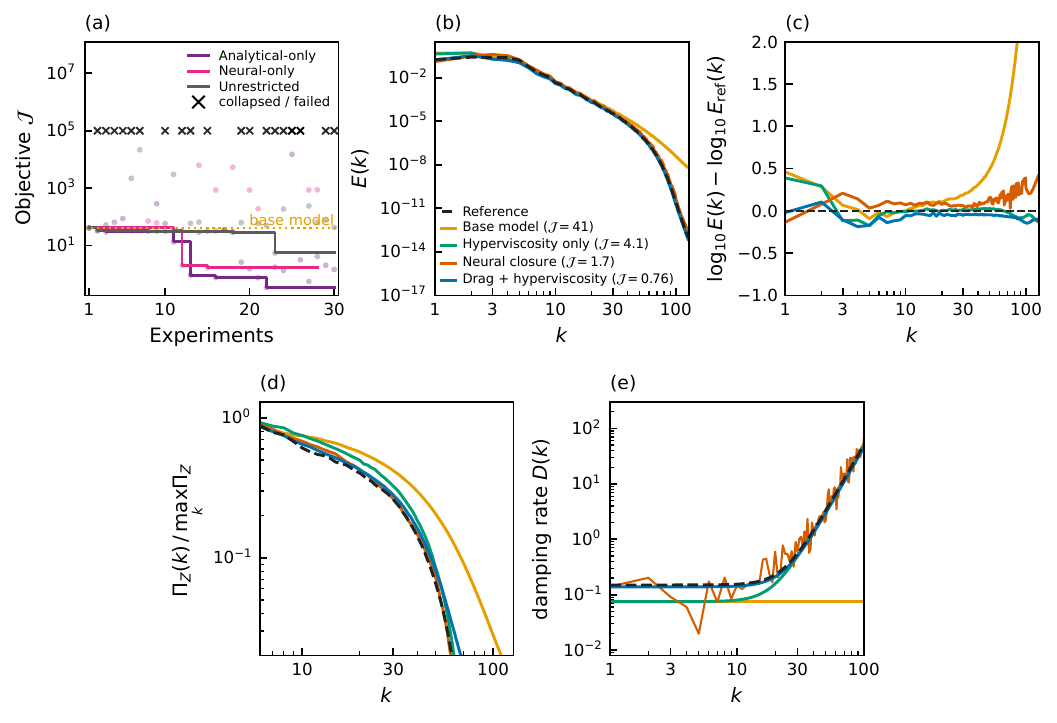}
  \caption{(a) Case~1 discovery trajectories from three runs (analytical-only, neural-only, and unrestricted): dots, experiments; step lines, best objective to date; crosses, rejected or failed experiments. (b--e) Comparison of three representative discovered closures of Case~1 with the base model and the reference in terms of (b) energy spectra, (c) log-spectrum error, (d) normalized enstrophy flux, and (e) damping rate \(D(k)\).}
  \label{fig:case1-overview}
\end{figure}

The first case concerns the discovery of a small-scale dissipation mechanism in forced two-dimensional turbulence, a setting with sufficiently low computational cost to permit extensive iteration. The flow is simulated on a \(256^2\) periodic grid over \([0,2\pi]^2\) with viscosity \(\nu = 8\times10^{-4}\) and driven by the Kolmogorov force \(\bm{f}_K(x,y) = (A\sin(k_f y),\,0)\), with amplitude \(A = 1.5\) and wavenumber \(k_f = 5\). Following the use of separate large-scale energy and small-scale enstrophy sinks in forced two-dimensional turbulence \citep{maltrud1991,borue1993,boffetta2007,boffetta2012}, the reference flow combines linear drag with fourth-order hyperviscosity. In Fourier space the dissipation force \(\bm{f}_d\) is
\begin{equation}
\hat{\bm{f}}_d(\bm{k}) \;=\; -\left(\alpha + \nu_h k^4\right)\hat{\bm{u}}(\bm{k}),
\qquad \alpha = 0.15, \quad \nu_h = 5\times10^{-7},
\label{eq:two-sink}
\end{equation}
where \(\widehat{(\cdot)}\) denotes the Fourier transform, \(\bm{k}\) is the wavevector, and \(k = |\bm{k}|\). This choice is motivated by the inverse transfer of energy and the forward transfer of enstrophy in two-dimensional turbulence \citep{kraichnan1967,batchelor1969}. Linear drag provides a large-scale energy sink, while fourth-order hyperviscosity provides a small-scale enstrophy sink, steepening the high-wavenumber spectrum. The baseline model retains the same viscosity and numerical settings but omits hyperviscosity and uses linear drag with \(\alpha = 0.075\), half the reference value. It therefore differs from the reference in both large- and small-scale damping. The agent is asked to match the reference kinetic-energy spectrum using the objective
\begin{equation}
\mathcal{J} \;=\; 100 \sum_{k=1}^{100} w(k)\,\big(\log_{10}E(k) - \log_{10}E_{\mathrm{ref}}(k)\big)^2,
\label{eq:case1-score}
\end{equation}
where \(E(k)\) and \(E_{\mathrm{ref}}\) are the simulated and reference energy spectra, respectively. The weights \(w(k) \propto 1/k\) are normalized to unit sum, giving approximately equal weight to each octave of scales. The agent may change the forcing parameters and replace the dissipation model with any force computed from the resolved velocity; the grid, viscosity, and numerical settings remain fixed. The closure therefore acts as a body force throughout the domain,
\begin{equation}
\mathcal{M}(\bm{u};\bm{\theta}) = \bm{f}_d(\bm{u};\bm{\theta}),
\label{eq:closure-force}
\end{equation}
with no prescribed structural template: the form, resolved inputs, and parameters are all free. The agent receives the reference spectrum, but not the dissipation law or coefficients used to generate it. The baseline gives \(\mathcal{J} = 41.4\), with spectral errors ranging from excess energy at \(k=1\) to an excess of more than three orders of magnitude at \(k=100\) (Fig.~\ref{fig:case1-overview}c). Because the reference dissipation has a compact analytical form within the agent's search space, this case also tests whether the search recovers the exact law or identifies an alternative that matches the spectrum.

Each run is fully autonomous under one of three closure-family regimes: \emph{analytical-only} (explicit analytical forms), \emph{neural-only} (trainable neural networks), or \emph{unrestricted} (all families admissible). A run consists of 30 sequential experiments, each proposed from the record of all previous ones. Figure~\ref{fig:case1-overview}a shows one run per regime: each dot is the objective of a single experiment, and the step line tracks the best objective to date. We examine three representative closures from these runs below, labeled as in Fig.~\ref{fig:case1-overview}: drag + hyperviscosity, hyperviscosity only, and the neural closure.

The analytical-only run recovers the form of the reference dissipation law. It adds a spectral hyperviscosity term to the base configuration and trains its coefficient together with the drag \emph{a posteriori} through the solver. The resulting closure has the same form as Eq.~\eqref{eq:two-sink}, with trained \(\alpha = 0.138\) and \(\nu_h = 4.5\times10^{-7}\) compared with reference values of \(0.15\) and \(5\times10^{-7}\), and achieves \(\mathcal{J} = 0.76\). The run's final experiment adds a Smagorinsky eddy-viscosity term to this closure, lowering the objective to \(0.34\) while moving \(\nu_h\) away from the reference value.

Another proposed analytical closure adds hyperviscosity to the otherwise unmodified base configuration and trains its single coefficient, yielding \(\nu_h = 5.2\times10^{-7}\) and \(\mathcal{J} = 4.06\). The residual error is concentrated at \(k=1\), where the drag mismatch remains uncorrected. That \(\nu_h\) is recovered closely without correcting the drag indicates that the high-wavenumber spectrum constrains the small-scale sink independently of the large-scale one.

The agent also proposes a spectral neural closure: a multilayer perceptron (MLP) with \(2.6\times10^4\) parameters maps the instantaneous log energy spectrum \(\log E(k)\) to a non-negative coefficient \(\alpha_{\rm NN}(k)\), applied as \(\hat{\bm{f}}_d(\bm{k}) = -\alpha_{\rm NN}(k)\,k^2\,\hat{\bm{u}}(\bm{k})\). The trained network reproduces the broad shape of the reference damping law (Fig.~\ref{fig:case1-overview}e), with approximately constant damping at large scales and a steep increase through the enstrophy range. It also exhibits a local dip at the forcing wavenumber \(k_f = 5\), which is absent from the reference law.
Although the coefficient depends on the instantaneous spectrum, its relative variation across the examined snapshots is below \(10^{-7}\), indicating that the trained network is effectively insensitive to its input. The network thus behaves as a fixed damping law in these simulations. With \(2.6\times10^4\) parameters trained \emph{a posteriori} through the chaotic rollout, it achieves \(\mathcal{J} = 1.67\).

\begin{figure}[!htp]
  \centering
  \includegraphics[width=\textwidth]{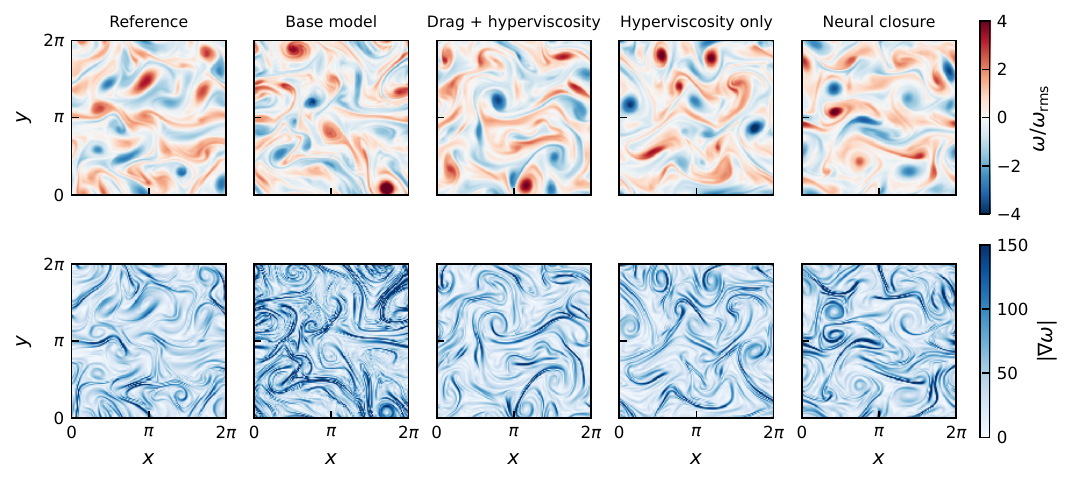}
    \caption{Instantaneous fields for Case~1 in the statistically stationary state, for the reference, the base model, and three agent-discovered closures. Top: vorticity normalized by its RMS. Bottom: vorticity-gradient magnitude. Snapshots are from independent realizations, so individual structures are not expected to coincide across panels.}
  \label{fig:case1-fields}
\end{figure}

Figures~\ref{fig:case1-overview} and \ref{fig:case1-fields} compare the three closures with the base model. Although any force was admissible, all three take the form of spectral damping, \(\hat{\bm{f}}_d(\bm{k}) = -D(k)\,\hat{\bm{u}}(\bm{k})\), differing only in the rate \(D(k)\) at which each damps the velocity mode at wavenumber \(k\). Figure~\ref{fig:case1-overview}e compares these rates with the reference law. All three closely follow the reference spectrum across the wavenumber range (Fig.~\ref{fig:case1-overview}b,c). Their effect on interscale transfer is quantified by the enstrophy flux \(\Pi_Z(k) = \sum_{|\bm{k}'|\leq k} \mathrm{Re}\big[\hat{\omega}^*(\bm{k}')\,\widehat{(\bm{u}\cdot\nabla\omega)}(\bm{k}')\big]\), the rate of nonlinear enstrophy transfer across wavenumber \(k\) \citep{kraichnan1967,boffetta2007,boffetta2012}, with \(\omega\) the vorticity (Fig.~\ref{fig:case1-overview}d). In the reference, the hyperviscous sink halts the enstrophy cascade well before the grid scale, and the three closures do so at the same wavenumber; in the base model the cascade continues to the smallest resolved scales. The physical-space fields show consistent behavior (Fig.~\ref{fig:case1-fields}): the large-scale vorticity is similar across cases, but dense, sharp vorticity-gradient filaments appear in the base model and are absent from both the reference and the discovered closures.

\subsection{Case 2: SGS closure discovery in 2D turbulence}\label{case-2-coarse-grid-sgs-closure-discovery-in-2d-hit}

\begin{figure}[!tp]
  \centering
  \includegraphics[width=0.85\textwidth]{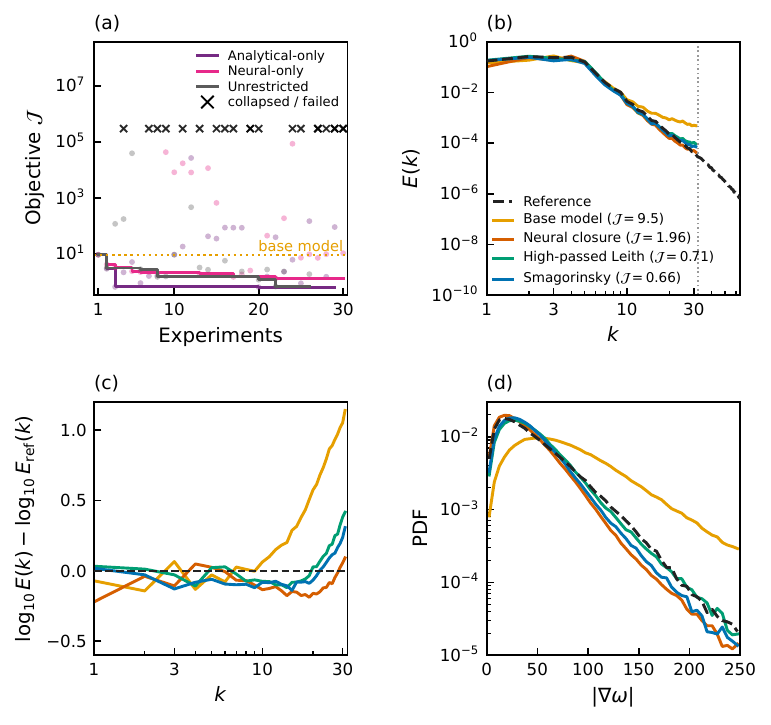}
  \caption{(a) Case~2 discovery trajectories from three runs (analytical-only, neural-only, and unrestricted): dots, experiments; step lines, best objective to date; crosses, rejected or failed experiments. (b--d) Comparison of three representative discovered closures of Case~2 with the base model and the DNS reference in terms of (b) energy spectra, (c) log-spectrum error, and (d) probability density function (PDF) of the vorticity-gradient magnitude.}
  \label{fig:case2-overview}
\end{figure}

The second case considers SGS modeling in forced two-dimensional turbulence with the same domain, viscosity, and Kolmogorov forcing as Case~1. A resolved \(512^2\) simulation provides the reference, while the base model uses the same physics on a \(64^2\) grid without an SGS closure. The coarse grid has a Nyquist wavenumber of \(k_c = N/2 = 32\) in each coordinate direction, limiting the resolved enstrophy cascade. The agent supplies a trainable SGS closure through the solver's eddy-viscosity interface,
\begin{equation}
\mathcal{M} = \nabla\cdot\big(2\nu_t(\bm{u};\bm{\theta})\,\bm{S}\big),
\label{eq:closure-sgs}
\end{equation}
where \(\bm{S}=\tfrac12(\nabla\bm{u}+(\nabla\bm{u})^{\mathsf T})\) is the strain rate tensor. The interface fixes the eddy-viscosity representation but leaves the dependence of \(\nu_t(\bm{x},t)\) on the resolved flow to the agent. The objective is Eq.~\eqref{eq:case1-score}, restricted to resolved wavenumbers \(k \in [1, 30]\). The base model gives \(\mathcal{J} = 9.5\), with energy accumulation near the cutoff and a tenfold excess at \(k = 30\) (Fig.~\ref{fig:case2-overview}c). Unlike Case~1, this case has no exact closure to recover. It therefore tests which forms the agent identifies and how strongly the objective distinguishes them.

Figure~\ref{fig:case2-overview}a shows one run per regime, as in Case~1. The analytical-only and unrestricted runs both reach objectives close to the noise level (\(\mathcal{J}\approx0.5\)), the former within the first few experiments and the latter after exploring several closure families. We examine three closures selected from these runs: the best closure of the analytical-only run (Smagorinsky), the best closure of the unrestricted run (a modified Leith model), and the best closure of the neural-only run (convolutional).

Within the eddy-viscosity interface, the analytical forms proposed by the agent are predominantly classical SGS models, which it then modifies and trains \emph{a posteriori} through the solver. The best closure of the analytical-only run is the Smagorinsky model, \(\nu_t = (C_s\Delta)^2|\bm{S}|\) \citep{smagorinsky1963}, where \(\Delta\) is the grid spacing and \(|\bm{S}|=\sqrt{2\bm{S}:\bm{S}}\). Its trained coefficient \(C_s = 0.154\) gives \(\mathcal{J} = 0.66\) with a mean eddy viscosity \(\langle\nu_t\rangle = 1.4\nu\). The trained \(C_s\) lies below Lilly's three-dimensional estimate of \(0.17\) \citep{lilly1967} and above the semi-analytical estimates of \(C_s = 0.11\)--\(0.13\) reported by \citet{guan2025} for a two-dimensional configuration with similar forcing and cutoff scales. The best closure of the unrestricted run builds on a Leith eddy viscosity \(\nu_t \propto |\nabla\omega|\) \citep{leith1968}, to which the agent adds a smooth high-pass filter of the eddy-viscosity field with cutoff wavenumber \(k = 12\) and a uniform offset equal to \(\nu\). Written in the standard form \(\nu_t = (C_L\Delta)^3|\nabla\omega|\), the trained amplitude corresponds to \(C_L = 0.35\), compared with \(C_L = 0.21\)--\(0.24\) in \citet{guan2025}. This comparison is approximate: the high-pass filter suppresses the large-scale contribution of the Leith term, and the offset supplies most of the dissipation (\(\nu\) of \(\langle\nu_t\rangle = 1.2\nu\)), so the closure acts predominantly as a constant eddy viscosity with a scale-selective correction. It gives \(\mathcal{J} = 0.71\). The two closures, found independently by different runs, are structurally different, yet their objective values differ by less than the noise level. The spectral objective therefore provides limited discrimination between these forms; distinguishing them may require additional statistics, such as the vorticity-gradient distribution in Fig.~\ref{fig:case2-overview}d, or conditions spanning several resolutions.

The best closure of the neural-only run is a local convolutional network with \(3\times3\) kernels and \(1.5\times10^4\) parameters. It maps four resolved scalar inputs (\(|\bm{S}|\), \(|\omega|\), \(|\bm{u}|^2\), and \(|\nabla^2\bm{u}|^2\)), similar to those of earlier learned SGS closures \citep{maulik2019,guan2022}, to a non-negative \(\nu_t(\bm{x})\). Trained \emph{a posteriori} through the chaotic rollout, it achieves \(\mathcal{J} = 1.96\) with \(\langle\nu_t\rangle = 2.8\nu\), twice the mean eddy viscosity of the Smagorinsky closure. Its spectrum lies slightly below the reference (Fig.~\ref{fig:case2-overview}c), consistent with excess dissipation.

\begin{figure}[!tp]
  \centering
  \includegraphics[width=\textwidth]{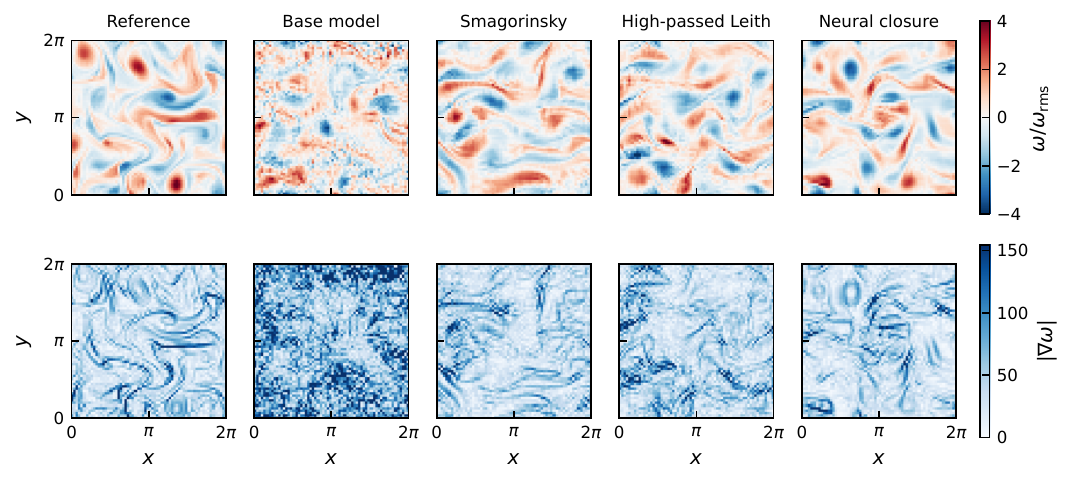}
  \caption{Instantaneous fields for Case~2, for the reference, the base model, and three agent-discovered closures. Top: vorticity normalized by its RMS. Bottom: vorticity-gradient magnitude. Snapshots are from independent realizations, so individual structures are not expected to coincide across panels.}
  \label{fig:case2-fields}
\end{figure}
Figures~\ref{fig:case2-overview} and \ref{fig:case2-fields} compare the three closures with the base model. All three suppress most of the energy accumulation and follow the reference spectrum up to \(k\approx25\), with a residual excess near the cutoff (Fig.~\ref{fig:case2-overview}b,c). In the base model, excess enstrophy appears as grid-scale noise in the vorticity and its gradient (Fig.~\ref{fig:case2-fields}) and as a vorticity-gradient probability-density tail approximately an order of magnitude above the reference (Fig.~\ref{fig:case2-overview}d). The discovered closures bring this tail close to the reference. The modified Leith closure agrees most closely, although its spectral loss is nearly equal to that of the Smagorinsky closure. The vorticity-gradient distribution thus distinguishes two closures that the spectral objective does not.

\paragraph{Transfer to unseen conditions}
We further evaluate the three discovered SGS closures beyond the training condition, keeping their model forms and trained parameters fixed, including the cutoff wavenumber and offset of the modified Leith closure. At the original forcing amplitude \(A=1.5\), the grid is changed from \(64^2\) to \(128^2\) and \(32^2\). At the original \(64^2\) resolution, the forcing amplitude is halved to \(A=0.75\) or doubled to \(A=3.0\). No discovery or retraining is performed at these conditions, and each case is compared with a \(512^2\) reference at the same forcing amplitude.

\begin{figure}[htp!]
  \centering
  \includegraphics[width=\textwidth]{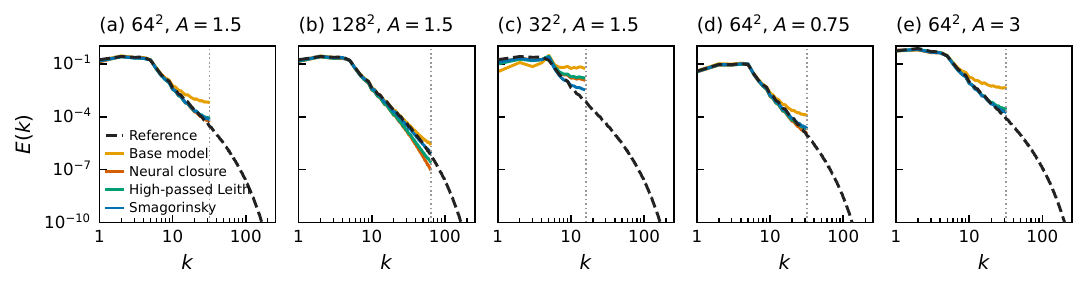}
    \caption{Energy spectra of the frozen Case~2 SGS closures at (a) the training condition (\(64^2\), \(A=1.5\)), (b) a finer grid (\(128^2\)), (c) a coarser grid (\(32^2\)), (d) half the training forcing amplitude (\(A=0.75\)), and (e) twice the training forcing amplitude (\(A=3.0\)). All closure parameters remain fixed at their trained values.}
  \label{fig:case2-transfer}
\end{figure}
Figure~\ref{fig:case2-transfer} shows that all three closures retain close agreement with the reference under changes in forcing amplitude, suppressing the excess small-scale energy observed without an SGS closure. Changing the grid resolution leads to larger differences. On the finer grid, the Smagorinsky closure continues to follow the reference closely, while the modified Leith and neural closures reproduce the overall spectral shape with slightly lower energy at high resolved wavenumbers. On the coarser grid, all three reduce the energy accumulation near the cutoff, again with the Smagorinsky closure in closest agreement. This difference follows from the closure forms: the Smagorinsky closure scales with the grid spacing by construction, whereas the fixed cutoff wavenumber and uniform offset of the modified Leith closure, and the inputs of the convolutional closure, do not adapt to resolution. Further improvements could be explored by combining objectives spanning multiple resolutions with physically motivated constraints on dimensional consistency and filter-width dependence \citep{chen2026physminer}.

\subsection{Case 3: wall-model discovery for 3D wall-bounded turbulence}\label{case-3-wall-model-discovery-for-3d-channel-flow}

\begin{figure}[htp!]
  \centering
  \includegraphics[width=0.65\textwidth]{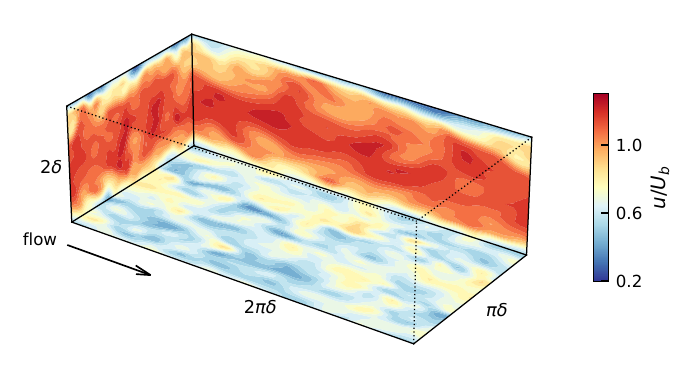}
  \caption{Instantaneous streamwise velocity \(u/U_b\) in the WMLES of turbulent channel flow (Case~3) with the hybrid convolutional closure, the agent-discovered wall model with the lowest objective value (described later in this section).}
  \label{fig:case3-flow}
\end{figure}

\begin{figure}[htp!]
  \centering
  \includegraphics[width=\textwidth]{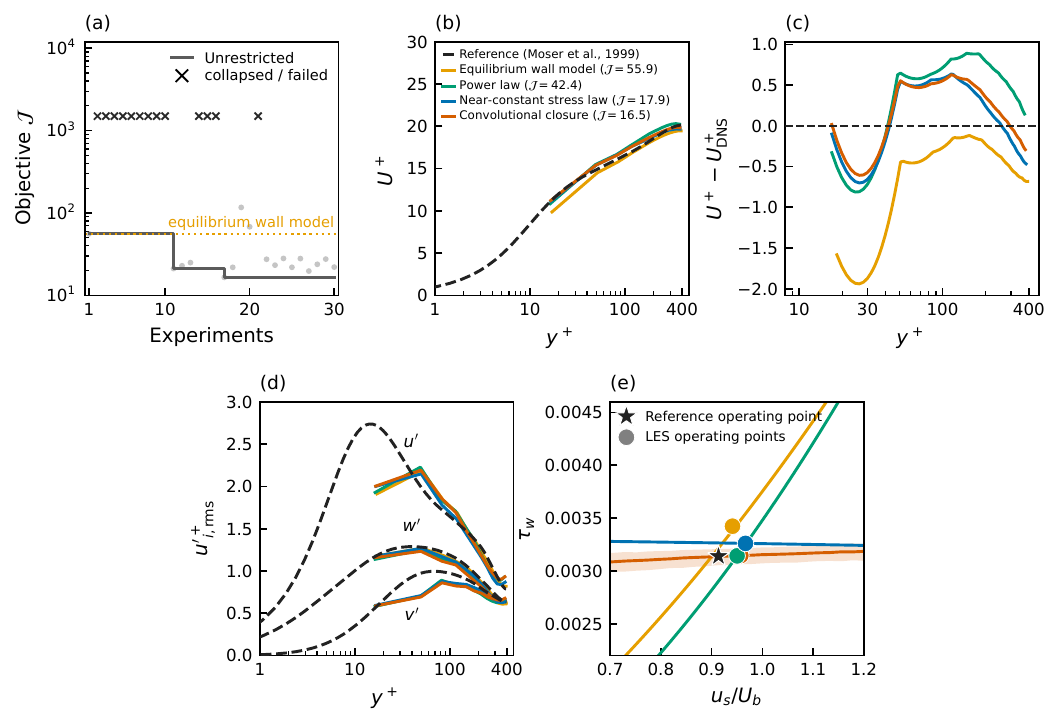}
  \caption{(a) Discovery trajectory for an unrestricted Case~3 run: dots, experiments; step line, best objective to date; crosses, experiments with wall-stress collapse, divergence, or other failures. (b--e) Comparison of three agent-discovered closures (power law, near-constant stress law, and hybrid convolutional closure) with the equilibrium wall model and the DNS reference in terms of (b) mean streamwise velocity profiles, (c) profile error, (d) root-mean-square velocity fluctuations, and (e) wall stress as a function of the matching-cell velocity, with the mean operating points (filled symbols) and the DNS operating point (star); for the hybrid closure the curve is the conditional median of its applied stress, with shading indicating the 5th--95th percentile range.}
  \label{fig:case3-overview}
\end{figure}

The third case extends the framework to three-dimensional wall-bounded turbulence, where the task is to discover a wall model for WMLES of turbulent channel flow (Fig.~\ref{fig:case3-flow}). The channel has half-height \(\delta = 1\) in a \(2\pi\times2\times\pi\) domain, resolved on a uniform \(48\times24\times48\) grid (\(\Delta x^+\approx51\), \(\Delta y^+\approx33\), \(\Delta z^+\approx26\)) with viscosity \(\nu = 1/7000\) and the Vreman SGS model \citep{vreman2004}. The reference is the DNS of \citet{moser1999} at friction Reynolds number \(Re_\tau = 392\). Here the closure enters through the boundary condition of Eq.~\eqref{eq:closure-bc}: the prescribed quantity \(\mathcal{B}\) is the wall shear stress \(\bm{\tau}_w\), representing momentum flux through the wall. The closure computes its magnitude from the resolved wall-parallel velocity \(\bm{u}_s\) at a matching height \(h_m\); the prescribed stress, taken as that exerted by the fluid on the wall, is aligned with this velocity,
\begin{equation}
\mathcal{M}_\Gamma(\bm{u};\bm{\theta}) = \bm{\tau}_w = \tau_w\,\frac{\bm{u}_s}{|\bm{u}_s|},
\qquad
\tau_w(\bm{x}_\parallel,t) = \tau_w\big(\bm{u}_s(\bm{x}_\parallel,t),\ldots;\bm{\theta}\big).
\label{eq:closure-wall}
\end{equation}
Following \citet{kawai2012}, the matching height is placed at the third off-wall cell center, \(h_m = 0.21\delta\) (\(h_m^+\approx82\)), away from the first cells where the resolved velocity is least accurate. The agent may construct additional inputs from the resolved flow, such as neighboring velocities and/or velocity gradients. The channel is driven at constant flow rate, with bulk velocity fixed to the DNS value; the friction velocity therefore emerges from the wall stress produced by the closure. The objective,
\begin{equation}
\mathcal{J} = 100\,\big\langle (U^+ - U^+_{\mathrm{DNS}})^2 \big\rangle_{y^+\geq10},
\label{eq:case3-score}
\end{equation}
where \(U^+ = \langle u\rangle/u_\tau\) and \(y^+ = y u_\tau/\nu\), with \(\langle\cdot\rangle\) denoting averaging over the homogeneous directions and time and \(u_\tau=\sqrt{\tau_w/\rho}\) obtained from the mean wall stress produced by each closure. The average in Eq.~\eqref{eq:case3-score} is taken over the LES grid points with \(y^+\geq10\), with the DNS profile interpolated to these points. Because both \(U^+\) and \(y^+\) are scaled by the closure's own friction velocity, \(\mathcal{J}\) is sensitive to both friction and profile-shape errors. The research question provides the DNS profile and \(Re_\tau\), which together with the bulk velocity determine the reference friction velocity; the agent thus knows the target friction level but must produce it through the closure. Proposals must depend on the resolved flow and return an explicit differentiable stress magnitude. The equilibrium wall model gives \(\mathcal{J} = 55.9\), overpredicting the friction (\(Re_\tau = 410\)) and shifting the velocity profile downward in wall units (Fig.~\ref{fig:case3-overview}b). This model is well suited to channel flow and provides a strong baseline despite the residual log-layer mismatch. The case is therefore intended to assess whether the discovery loop can operate autonomously in three-dimensional wall-bounded turbulence and produce a wall model of comparable performance, rather than to establish a superior wall model. To our knowledge, LLM-driven closure discovery has not previously been demonstrated in three-dimensional wall-bounded turbulence within a fully autonomous loop that includes implementation and training.

The unrestricted run (Fig.~\ref{fig:case3-overview}a) shows how numerical failures shape the search. Initially, many of its 30 experiments are rejected, become unstable, or cause the wall stress to collapse (\(u_\tau \to 0\)). The agent responds by converging on a common stabilization structure: a fixed anchor equal to the mean wall stress (or friction velocity) of the equilibrium baseline, supplemented by a trainable flow-dependent correction, which the agent chose to bound in its leading proposals. The anchor keeps the stress near a physically reasonable level, and the bound limits the risk of collapse during parameter updates; the framework itself imposes neither. The recorded hypotheses explicitly identify collapse prevention as the motivation for this construction. This sequence illustrates how the proposer adapts the closure parameterization in response to numerical failures, making further gradient-based training possible.

Both leading closures adopt this structure. The best analytical closure is a near-constant stress law, \( \tau_w = \tau_a + \Delta\tau\,\tanh\!\big(c_1 \tilde{u} + c_2 \tilde{u}^{\,2}\big) \), where \(\tilde{u} = u_s/U_b\) is the matching-cell streamwise velocity normalized by the bulk velocity \(U_b\); with \(\Delta\tau\), \(c_1\), and \(c_2\) trained \emph{a posteriori}, it achieves \(\mathcal{J} = 17.9\). The proposal with the lowest objective value overall is a hybrid closure in which a \(3\times3\) convolutional network over five wall-plane fields, with 3{,}202 parameters, supplies a bounded correction to the friction-velocity anchor, with \(\tau_w=\rho u_\tau^2\); it achieves \(\mathcal{J} = 16.5\). Both bring the friction Reynolds number closer to the DNS value of 392: \(Re_\tau = 400\) for the near-constant stress law and 397 for the hybrid closure, compared with 410 for the equilibrium wall model. Because the anchor is fixed at the equilibrium-model level, this 5\% reduction in mean wall stress is produced by the correction via \emph{a posteriori} training.

The residual error of the equilibrium wall model has two parts, the level of the mean wall stress and the shape of the velocity profile, and Fig.~\ref{fig:case3-overview} separates them. Both are visible in Fig.~\ref{fig:case3-overview}b,c: the overprediction of friction shifts the profile downward in wall units, and a shape error remains in the first off-wall cells. In the stress--velocity comparison (Fig.~\ref{fig:case3-overview}e), the stress--velocity relation of the equilibrium wall model, which inverts Spalding's law \citep{spalding1961} at the matching height, passes through the DNS operating point, but the coarse WMLES supplies a higher matching-cell velocity and consequently obtains a higher stress. This behavior is consistent with the established sensitivity of wall models to the matching location and the accuracy of the resolved input \citep{kawai2012}. An agent-proposed power law, \(\tau_w \propto u_s^2\), illustrates this distinction: coefficient adjustment recovers the reference friction, yet the objective remains \(\mathcal{J} = 42.4\). Correcting the mean stress level alone therefore does not account for the loss reduction achieved by the leading closures.

The near-constant stress law exhibits a weak dependence of stress magnitude on instantaneous matching velocity, and the hybrid neural closure learns a similarly flat response despite access to additional wall-plane information (Fig.~\ref{fig:case3-overview}e). Their lower values of the chosen objective are consistent with the benefit of reducing instantaneous stress--velocity coupling reported in input-filtered wall modeling \citep{yang2017}. The near-constant magnitude is also closely related to the mean-wall-shear-stress boundary condition of \citet{lee2013}, with the difference that the mean stress here is not prescribed but obtained through \emph{a posteriori} training from the equilibrium-model level. Physically, the mean streamwise wall stress sets the mean momentum flux that the LES must transport; reproducing that flux does not require reproducing every fluctuation of the wall stress. The present results suggest that a weakly varying stress magnitude can provide an effective boundary condition for this coarse channel-flow calculation. The root-mean-square velocity fluctuation profiles, which are not included in the objective, remain similar for the equilibrium wall model and the three discovered closures (Fig.~\ref{fig:case3-overview}d).

The loop thus operates autonomously in three-dimensional wall-bounded turbulence and produces wall models with performance comparable to the equilibrium baseline and stress responses consistent with established wall-modeling approaches. The discovered closures attain lower values of the chosen objective, while the equilibrium model retains closer agreement with the reference profile in the logarithmic region (Fig.~\ref{fig:case3-overview}b,c). Because the leading closures are trained at a single Reynolds number with a near-constant stress magnitude, their transfer across flow conditions remains to be assessed.

\subsection{Patterns in closure proposals across runs}\label{proposals-across-runs}

The preceding sections examine representative closures. The complete set of proposals, including failed experiments, shows how the agent uses established closure theory and standard neural architectures as starting points and develops them through simulation feedback. For analytical proposals, these starting points are primarily canonical turbulence closures. For dissipation-law discovery, they include linear drag, hyperviscosity, large-scale drag restricted to the lowest wavenumbers, spectral eddy viscosity \citep{kraichnan1976}, and eddy viscosities based on the strain rate or vorticity. For SGS modeling, they include the Smagorinsky and Leith models \citep{smagorinsky1963,leith1968} and strain-dependent coefficients related to the dynamic procedure \citep{germano1991}; the agent also proposed scale-similarity \citep{bardina1980}, mixed similarity--viscosity, and stochastic backscatter \citep{leith1990} models. Analytical wall-model proposals include power-law and logarithmic wall functions with adjustable constants.

The neural proposals use standard architectures to represent closure quantities such as damping rates, eddy viscosity, and wall stress. They include pointwise networks acting on local flow features, convolutional networks acting on spatial neighborhoods of the resolved fields, and networks that predict damping in spectral space. For these candidates, the agent selects the inputs and network structure, while the dependence of the closure output on those inputs is learned through \emph{a posteriori} training.

Building on these starting points, the agent revises closure forms and inputs and combines analytical and learned components in response to simulation results. In Case~1, it concentrated dissipation at the wavenumbers most strongly weighted by the objective and trained a neural damping law without imposing the analytical two-sink form. In Case~2, it explored spectral tapering and high-pass filtering of the eddy-viscosity field to target the coarse-grid cutoff, and made the Smagorinsky coefficient increase with local strain relative to its mean. It also conditioned a convolutional closure on a global descriptor of the resolved spectrum through feature-wise linear modulation (FiLM) \citep{perez2018}, giving the local network access to information about the energy cascade. In Case~3, it introduced a shear estimate from two wall-normal cells and stencil averages of the wall-plane velocity as inputs, and developed the anchored, bounded-correction formulation described in Section~\ref{case-3-wall-model-discovery-for-3d-channel-flow}. These modifications emerged sequentially within each run, as the agent responded to the limitations and failures of earlier proposals, rather than from a predefined closure template.

The search thus uses established models as a basis for problem-dependent modifications and combinations. More recent or specialized turbulence closures do not appear among the identifiable starting points in the reported runs, consistent with proposals drawn from the modeling knowledge most strongly represented in the LLM's training data. Literature retrieval could broaden the physical models and neural modeling strategies on which the agent builds its proposals.

\subsection{Role of gradients: solver modes and feedback ablation}\label{ablation-of-the-feedback-modes}
\begin{figure}[!tp]
  \centering
  \includegraphics[width=0.92\textwidth]{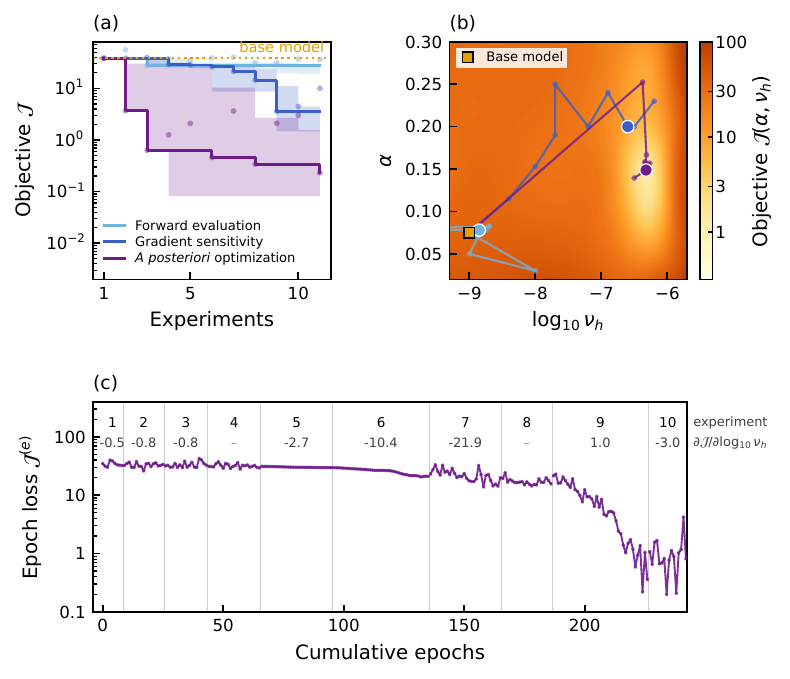}
  \caption{Comparison of solver modes. (a) Discovery trajectories, three runs per mode (line, median; band, range; dotted, base model). (b) Parameter trajectories for one representative run per mode, overlaid on the objective evaluated on a coefficient grid (lighter shading indicates lower values; square, base model; large circles, best experiments). (c) Loss by epoch for the optimization run used in the gradient-feedback ablation. Labels above the plot indicate experiment numbers and final reported gradients; blanks indicate experiments with fixed hyperviscosity.}
  \label{fig:ladder}
\end{figure}

In the discovery runs of Sections~\ref{case-1-dissipation-law-discovery-in-2d-hit}--\ref{case-3-wall-model-discovery-for-3d-channel-flow}, the agent selects solver modes freely and changes closure forms between experiments, so the contribution of gradients cannot be isolated. We isolate it here through two controlled tests: a comparison of solver modes with the closure form held fixed, and an ablation of the gradient reported to the proposer. For the first, the Case~1 problem is repeated with the two-sink law in Eq.~\eqref{eq:two-sink}, leaving only its two coefficients unknown, under the same budget of experiments for each mode. At initialization, the hyperviscosity is a factor of one thousand below the reference value and is therefore nearly inactive. Figure~\ref{fig:ladder}a summarizes three runs per mode, showing the median trajectory and the range, while Fig.~\ref{fig:ladder}b tracks a representative run in coefficient space. This initialization creates a challenging search: the objective varies weakly with hyperviscosity on the plateau in Fig.~\ref{fig:ladder}b, so objective values alone provide little guidance for adjusting this coefficient.

The three modes produce distinct search trajectories across this plateau. With forward evaluation alone, the agent probes small changes in both coefficients but observes little response to hyperviscosity and repeatedly retunes the drag. In gradient-sensitivity mode, the derivative signs guide parameter changes despite their small magnitudes: the agent adjusts one coefficient at a time, increasing hyperviscosity by factors of two to three until it reaches the low-objective region within the budget. In \emph{a posteriori} optimization, the solver takes repeated gradient steps within each experiment, and subsequent experiments resume from the best retained parameters; this sequence recovers the coefficients that generate the reference spectrum (Fig.~\ref{fig:ladder}b).

In optimization mode, gradients serve two roles: they drive parameter updates within each experiment and inform the proposer's choices for subsequent experiments. The proposer receives the training loss history together with the gradient from the lowest-loss epoch. We examine the second role through a standalone ablation of gradient feedback to the proposer, using the optimization run in Fig.~\ref{fig:ladder}c. Experiments 1--4 stalled on the plateau and terminated through early stopping; after each, the proposer reduced the learning rate, from \(0.1\) to \(0.002\). Experiment~5 again reduced the loss by less than the estimated noise level, but returned a gradient of \(-2.7\), compared with \(-0.5\) to \(-0.8\) on the plateau. The proposer interpreted the larger magnitude and unchanged sign as increasing sensitivity along the search direction, cited this gradient, and retained the learning rate; by experiment~9, the run had recovered the reference coefficients.

To isolate the effect of the reported gradient on the next proposal, we replayed this decision with the gradient removed from the experiment record, retaining the loss history and optimization results. The proposer continued adjusting hyperviscosity because the loss was decreasing, but lowered the learning rate in every replay. This ablation shows that gradient feedback can affect the proposer's choice of optimization settings even when the outcomes of gradient-based training are already available. At this step, the reported sensitivity supported retaining the learning rate, whereas the remaining feedback prompted more conservative updates.

These results illustrate how the agent uses gradients at both levels of the closure discovery. Within experiments, gradient-based optimization recovers coefficients that forward evaluation alone does not reach and makes training practical for closures with many parameters, including the neural and hybrid closures of Sections~\ref{case-1-dissipation-law-discovery-in-2d-hit}--\ref{case-3-wall-model-discovery-for-3d-channel-flow}. Across experiments, sensitivities guide the agent’s parameter search when changes in the objective are small, and reported gradients inform its subsequent proposals, including the choice of optimization settings. Other solver feedback complements these gradient signals: objective values and diagnostics guide revisions of closure forms throughout the runs, and numerical failures can prompt structural changes, as in the wall-model discovery study (Section~\ref{case-3-wall-model-discovery-for-3d-channel-flow}).

\section{Conclusions and outlook}\label{conclusions-and-outlook}

We have presented a gradient-aware agentic framework for turbulence model discovery that poses the problem as a bilevel search: LLM agents explore closure forms and physical inputs, while a differentiable CFD solver supplies gradients for \emph{a posteriori} training through simulations of chaotic flow. Differentiable \emph{a posteriori} learning thereby becomes an action of the agent. For each proposal, the agent constructs and validates a differentiable training program, selects among forward evaluation, gradient sensitivity, and \emph{a posteriori} optimization, and uses the returned gradients and outcomes to revise subsequent proposals. It also derives a differentiable objective from the research question when one is not specified, and selects warm-up, averaging, and differentiated horizons from the flow time scales to support training under chaotic dynamics.

The dissipation-law, SGS, and channel-flow wall-model cases demonstrate autonomous discovery and training of analytical, neural, and hybrid closures in time-resolved two- and three-dimensional turbulence, including networks with \(\mathcal{O}(10^4)\) parameters. With the closure form fixed, gradient-based training recovered reference coefficients that forward evaluation alone did not reach within the same budget, and in the wall-model study \emph{a posteriori} training brought the predicted friction closer to the DNS value than the equilibrium baseline. The experiments also show how differentiable solver feedback shapes the search: reported sensitivities changed the proposer's optimization settings, and numerical failures led the agent to an anchored, bounded-correction parameterization. Flexible representations sometimes reduced to simple effective behavior, such as nearly input-independent spectral damping and a weakly varying wall-stress magnitude.

These canonical flows provide a controlled setting for assessing the complete discovery cycle, from translating the research question into a differentiable objective through closure proposal, implementation, validation, \emph{a posteriori} training, and revision. The discovered closures largely build on established mechanisms and model forms; comparisons with these models make the results interpretable and help distinguish successful workflow execution from physical model validity \citep{wang2026closure}.

The present study also has limitations. The closures were developed from limited reference data, each targeting a single statistic in a single flow configuration, and the spectral objective did not discriminate between structurally different SGS closures. The transfer tests provide an initial assessment beyond the training conditions, but transfer across distinct flow configurations remains to be assessed. As reference data spanning different flows become available, future work could combine multiple target statistics with physical constraints and training across flows and resolutions, followed by independent testing without further tuning. The reported studies used GPT-5-mini throughout; the framework accommodates other LLMs, and more capable models may improve closure proposals.

LLM-driven proposals have also been applied to flows in which existing closures are less reliable. Working with DeepSeek-R1, \citet{yang2025} obtained wall models for adverse-pressure-gradient, rotating, and rough-wall flows, including a hybrid analytical--neural roughness law, that improved upon baseline models in \emph{a posteriori} tests, with human researchers implementing, training, and evaluating the candidates. Combining such proposals with the present autonomous implementation and training workflow, and equipping the proposer with literature retrieval as in \citet{somasekharan2026}, could extend the search beyond the canonical starting points observed in Section~\ref{proposals-across-runs} to flows involving pressure gradients, separation, rotation, and roughness. Because the framework already derives objectives and incorporates successful closures into the base for subsequent experiments, allowing the agent to also formulate new research questions from its results would extend it toward open-ended discovery with evolving objectives \citep{stanley2019,hughes2024,lu2026,somasekharan2026}.

The bilevel approach can apply beyond turbulence wherever a differentiable forward model is available, including subgrid parameterizations in combustion, multiphase flow, and climate modeling, and constitutive-model discovery problems in which parameter gradients are available but model-form selection remains manual.

\section*{Acknowledgements}
\label{}

The authors gratefully acknowledge funding from the U.S. Department of Energy, Office of Science, under award No.~DE-SC0026687 and award No.~DE-SC0026662, as part of the Genesis Mission: Transforming Science and Energy with AI.

\appendix

\section{Workflow graph}\label{app:workflow-graph}

This appendix presents the complete state graph summarized in Section~\ref{system-overview}. The simulation pipeline converts a flow description into a verified solver script and executes it. The auto-research loop for closure discovery in Section~\ref{closure-proposal-and-implementation} then proposes, implements, validates, trains, and analyzes closures until the experiment budget is exhausted. Conditional edges represent retry and repair paths, and review nodes identify optional human intervention points.

\begin{figure}[!ht]
  \centering
  \includegraphics[width=\textwidth]{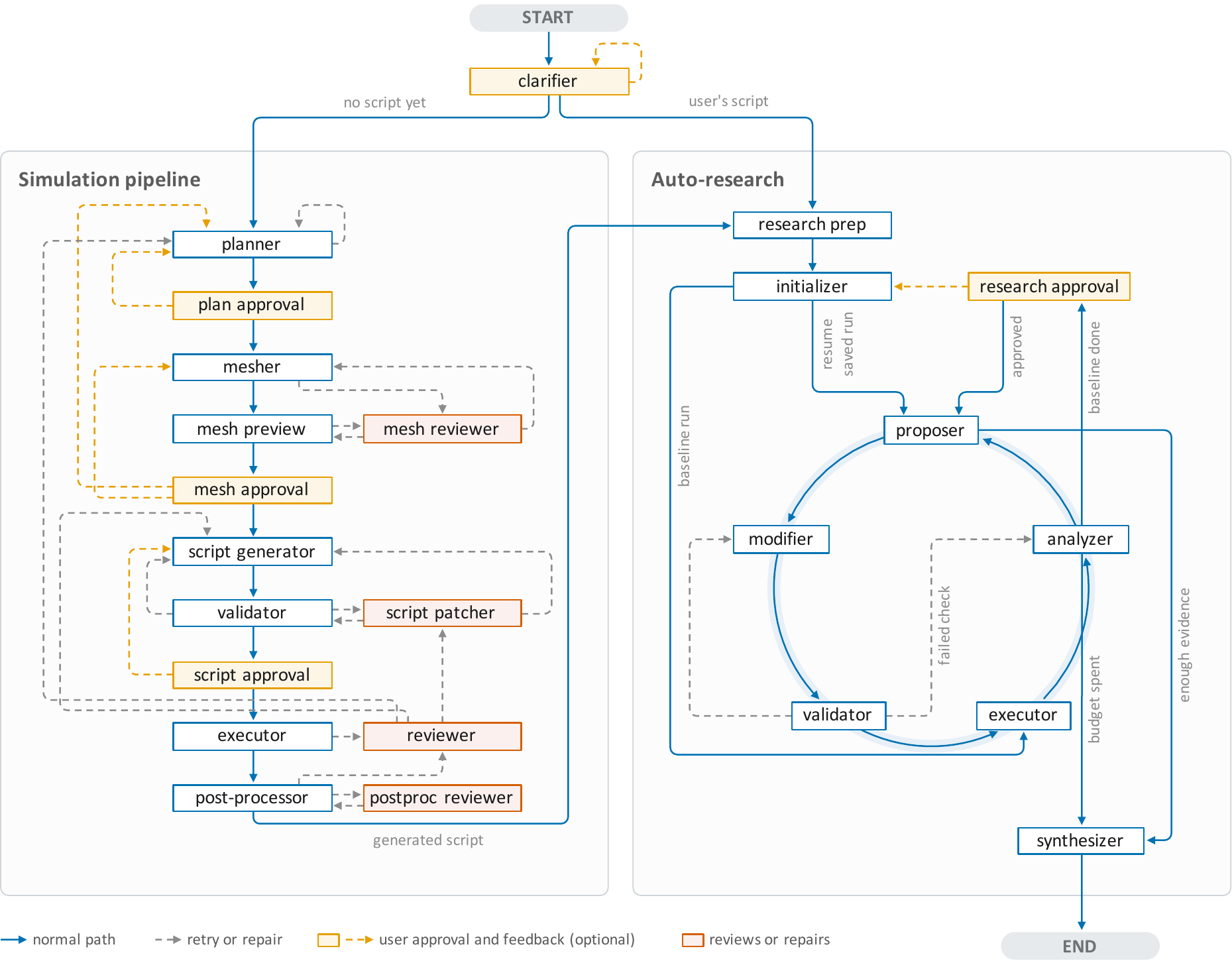}
  \caption{Node-and-edge structure of the agent workflow: the simulation pipeline that produces a verified solver script, and the auto-research loop that proposes, trains, and analyzes closures.}
  \label{fig:workflow-graph}
\end{figure}

\clearpage
\section{Transcript of a complete run}\label{app:chat-transcript}

This appendix reproduces a user--agent conversation for the problem in Section~\ref{case-1-dissipation-law-discovery-in-2d-hit}. The agent generates the base simulation from the user's prompt and pauses at each review point for user feedback, demonstrating the workflow from research question to final report and model. For brevity, three of the 30 experiments are shown in full, with longer passages abridged.

\chatframeon
\begin{usermsg}
Simulate 2D forced turbulence in a doubly periodic 2*pi box on a 256x256 grid with viscosity 8e-4, driven by a Kolmogorov forcing 1.5*sin(5*y) in x plus a linear friction of 0.075. Then improve the friction/dissipation model so that the energy spectrum matches the reference spectrum saved at reference\_spectrum.npz (keys 'k' and 'E'). Only change the forcing and dissipation model; no neural networks.
\end{usermsg}

\begin{agentmsg}
Running workflow\ldots
\stageheader[Sky]{Clarifier}{classifying intent}
\steprow{Clarifier}
\inforow[Sky]{Clarifier}{Intent: research\_combined (2.2s)}
\detailrow{Extracted research part: Improve the friction/dissipation model so that the energy spectrum matches the provided reference spectrum (only change forcing and dissipation model; no neural networks).}
\donerow{Clarifier}{research\_combined intent}
\transrow{Sky}{Clarifier}{Vio}{Planner}{simulate intent}
\stageheader[Vio]{Planner}{parsing user requirement}
\inforow[Vio]{Planner}{Loaded solver spec (39065 chars)}
\inforow[Vio]{Planner}{Calling LLM to generate simulation plan (\textasciitilde{}25,149 tokens)...}
\steprow{Planner}
\errrow{Plan validation failed with 2 error(s)}
\detailrow{Plan field 'external\_forces $\rightarrow$ 0': Input should be a valid string}
\detailrow{Plan field 'external\_forces $\rightarrow$ 1': Input should be a valid string}
\warnrow{Plan validation failed (attempt 1/10) --- will retry.}
\transrow{Vio}{Planner}{Vio}{Planner}{JSON parse failed $\rightarrow$ retry}
\stageheader[Vio]{Planner}{parsing user requirement}
\steprow{Planner}
\inforow[Vio]{Planner}{Plan saved to \path{agent_runs/KolmogorovBox_n256_nu8e-4/plan.json}}
\donerow{Planner}{}
\transrow{Vio}{Planner}{Amb}{Plan Approval}{plan ready for review}
\stageheader[Amb]{Plan Approval}{awaiting user review}
\renewcommand{\planaccent}{uiSuccess}
\begin{plancard}{Simulation Plan}
\plansec{Problem}
\planrow{Description}{Base case: 2D doubly-periodic Kolmogorov-forced turbulence (no IBM). Keep unmodified physics as the research BASE: single Fourier-mode Kolmogorov forcing f\_x = 1.5*sin(5*y) and linear friction alpha=0.075. The reference spectrum for later improvement is located at reference\_spectrum.npz (keys 'k' and 'E'). Derived quantities (literal substitutions shown): \elide{paragraphs not shown} Notes: this plan is the unmodified BASE. The user-requested improvement (modify forcing and dissipation to match the reference spectrum, no neural networks) should be implemented by the research agent by changing only the external\_forces and the dissipation model in subsequent experiments. The base configuration here uses a semi-implicit collocated solver, stable upwind collocated advection, and no SGS model (DNS-style). The external forcings specified: Kolmogorov forcing amplitude=1.5 at k\_y=5 producing f\_x=1.5*sin(5*y); linear friction coefficient alpha=0.075. ICs: random small perturbations (amplitude \textasciitilde{}1\% of U) are required when using any SGS model; base DNS still benefits from noise for spectral richness. Output folder: 'KolmogorovBox\_n256\_nu8e-4'.}
\planrow{Dimensionality}{2D}
\planrow{Grid type}{collocated}
\planrow{Problem type}{periodic\_kolmogorov\_forced\_turbulence}
\planrow{Simulation type}{DNS}
\plansec{Geometry}
\planrow{Domain}{Lx=6.283185307179586, Ly=6.283185307179586}
\planrow{Resolution}{nx=256, ny=256}
\planrow{Stretching}{disabled}
\plansec{Physics}
\planrow{Reynolds number}{7853.981633974483}
\planrow{Re definition}{direct\_viscosity}
\planrow{Density}{1.0}
\planrow{Viscosity}{0.0008}
\plansec{Boundary Conditions}
\planrow{BC type}{periodic}
\plansec{IBM}
\planrow{IBM enabled}{no}
\planrow{IBM interpolation method}{sharp}
\planrow{Compute forces}{no}
\plansec{Numerics}
\planrow{Solver type}{semi\_implicit\_navier\_stokes}
\planrow{Advection scheme}{advect\_upwind\_colocation}
\planrow{Pressure solver}{solve\_ncg}
\planrow{Subgrid model}{none}
\planrow{PETSc enabled}{no}
\planrow{x64 precision}{no}
\plansec{Time Stepping}
\planrow{dt}{0.0196349538385368}
\planrow{CFL safety}{0.8}
\planrow{Steps}{10 inner x 32 outer x 2 loops = 640 total}
\planrow{Loop / restart}{enabled (2 loops)}
\plansec{Output}
\planrow{Output folder}{KolmogorovBox\_n256\_nu8e-4}
\planrow{External forces}{kolmogorov\_forcing: amplitude=1.5, k=5, form=f\_x=1.5*sin(5*y) (x-directed), linear\_friction: alpha=0.075 (linear Rayleigh friction applied to velocity)}
\end{plancard}
\end{agentmsg}

\begin{agentmsgc}
Do you approve this simulation plan?\par\vspace{4pt}
\btnapprove{0}\hspace{4pt}\btnfeedback{1}{Give feedback (revise)}\hspace{4pt}\btnreject{0}
\end{agentmsgc}

\begin{agentmsgc}
What should be changed in the plan? (describe your feedback)
\end{agentmsgc}

\begin{usermsg}
Please use a staggered grid with central advection and the FFT pressure solver. Use dt 0.001 with 10 inner x 100 outer steps per loop, keep 2 loops with restart.
\end{usermsg}

\begin{agentmsg}
\stageheader[Vio]{Planner}{re-planning with user feedback}
\inforow[Vio]{Planner}{Incorporating user feedback: Please use a staggered grid with central advection and the FFT pressure solver. Use dt 0.001 with \elide{truncated}}
\steprow{Planner}
\inforow[Vio]{Planner}{Plan saved to \path{agent_runs/KolmogorovBox_n256_nu8e-4_staggered/plan.json}}
\donerow{Planner}{}
\transrow{Vio}{Planner}{Amb}{Plan Approval}{plan ready for review}
\stageheader[Amb]{Plan Approval}{awaiting user review}
\renewcommand{\planaccent}{uiSuccess}
\begin{plancard}{Simulation Plan}
\plansec{Problem}
\planrow{Description}{Base case: 2D doubly-periodic Kolmogorov-forced turbulence (no IBM). Keep unmodified physics as the research BASE: single Fourier-mode Kolmogorov forcing f\_x = 1.5*sin(5*y) and linear friction alpha=0.075. Reference spectrum for later improvement is located at reference\_spectrum.npz (keys 'k' and 'E'). \elide{paragraphs not shown} Notes and constraints: This plan is the unmodified BASE. The user will later improve forcing and dissipation (only those two components) to match the reference spectrum (no neural networks). The solver is set to staggered-grid semi-implicit Navier--Stokes with central (linear) advection as requested, and FFT-based periodic Poisson solver for pressure. No SGS model (DNS-style) is used here. Initial condition: small random perturbations (\textasciitilde{}1\% U) are recommended for spectral richness when running and mandatory if any SGS model is later enabled. Output folder: 'KolmogorovBox\_n256\_nu8e-4\_staggered'. Loop restart enabled with num\_loops=2 as requested. \elide{paragraphs not shown}}
\planrow{Grid type}{staggered}
\plansec{Numerics}
\planrow{Advection scheme}{advect\_linear}
\planrow{Pressure solver}{solve\_fast\_diag}
\planrow{Time stepper}{mixed\_scheme}
\plansec{Time Stepping}
\planrow{dt}{0.001}
\planrow{Steps}{10 inner x 100 outer x 2 loops = 2000 total}
\plansec{Output}
\planrow{Output folder}{KolmogorovBox\_n256\_nu8e-4\_staggered}
\planline{\elide{rows unchanged from the previous plan not shown}}
\end{plancard}
\inforow[Amb]{Approval}{Previous feedback was: Please use a staggered grid with central advection and the FFT pressure solver. Use dt 0.001 with 10 inner x 100 outer steps per loop, keep 2 loops with restart.}
\end{agentmsg}

\begin{agentmsgc}
Do you approve this simulation plan?\par\vspace{4pt}
\btnapprove{1}\hspace{4pt}\btnfeedback{0}{Give feedback (revise)}\hspace{4pt}\btnreject{0}
\end{agentmsgc}

\begin{agentmsgc}
\stageheader[Grn]{Mesher}{refining resolution \& geometry}
\steprow{Mesher}
\inforow[Grn]{Mesher}{Resolution: \{'nx': 256, 'ny': 256, 'nz': None\} $\rightarrow$ \{'nx': 256, 'ny': 256, 'nz': None\}}
\inforow[Grn]{Mesher}{Stretching: disabled}
\detailrow{Reasoning: Kept the user-specified resolution (nx=256, ny=256) unchanged as requested. This is a doubly-periodic 2D DNS Kolmogorov-forced box with Lx = Ly = 2$\pi$, giving dx = dy = 6.283185307179586 / 256 = 0.024543692298171. The spectral Nyquist/resolvable wavenumber is k\_max = pi / dx = 128.0. Rationale and scale estimates \elide{truncated}}
\inforow[Grn]{Mesher}{Mesher output saved to \path{agent_runs/KolmogorovBox_n256_nu8e-4_staggered/mesher_output.json}}
\donerow{Mesher}{}
\transrow{Grn}{Mesher}{Cya}{Mesh Preview}{mesh output OK $\rightarrow$ visualize}
\stageheader[Cya]{Mesh Preview}{generating mesh visualization}
\inforow[Cya]{MeshPreview}{2D mesh preview saved to \path{agent_runs/KolmogorovBox_n256_nu8e-4_staggered/mesh_preview.png}}
\donerow{Mesh Preview}{}
\transrow{Cya}{Mesh Preview}{Amb}{Mesh Approval}{preview ready for review}
\stageheader[Amb]{Mesh Approval}{awaiting user review}
\inforow[Amb]{Approval}{Mesh preview image: \path{agent_runs/KolmogorovBox_n256_nu8e-4_staggered/mesh_preview.png}}
\end{agentmsgc}

\begin{agentmsgc}
Mesh preview
\imgel{0.42\linewidth}{mesh_preview.png}
\end{agentmsgc}

\begin{agentmsgc}
Does the mesh / geometry look correct?\par\vspace{4pt}
\btnapprove{1}\hspace{4pt}\btnfeedback{0}{Give feedback (revise mesh)}\hspace{4pt}\btnreject{0}
\end{agentmsgc}

\begin{agentmsgc}
\stageheader[Blu]{Script Generator}{first generation}
\inforow[Blu]{ScriptGen}{Primary demo: 2D Homogeneous Isotropic Turbulence (staggered)}
\detailrow{Selection reason: exact match on dimensionality=2D, grid\_type=staggered, implicit solver family (feature score 1: \{'loop\_restart'\})}
\inforow[Blu]{ScriptGen}{Calling LLM to generate simulation script (\textasciitilde{}23,766 tokens input)...}
\steprow{ScriptGen}
\inforow[Blu]{ScriptGen}{Generation completed in 43.5s}
\inforow[Blu]{ScriptGen}{Generated script: 10817 chars, 297 lines}
\okrow{Script saved to \path{agent_runs/KolmogorovBox_n256_nu8e-4_staggered/simulation.py}}
\donerow{Script Generator}{}
\stageheader[Cya]{Validator}{static analysis}
\inforow[Cya]{Validator}{Checking script (10817 chars, 297 lines)}
\detailrow{Syntax OK $\cdot$ pyflakes: clean $\cdot$ Signature check: all calls OK $\cdot$ Attribute chains: all references valid}
\okrow{All validation checks passed}
\donerow{Validator}{PASSED}
\transrow{Cya}{Validator}{Amb}{Script Approval}{validation passed}
\stageheader[Amb]{Script Approval}{awaiting user review}
\begin{logblock}{Generated script}{python $\cdot$ 297 lines}
\cl{import os}
\cl{import time as ti}
\cl{import jax}
\cl{import jax.numpy as jnp}
\cl{import xarray}
\cl{import numpy as np}
\cl{import Jax\_FlowFSI.src as fsi}
\cl{output\_folder = \textquotedbl{}\path{agent_runs/KolmogorovBox_n256_nu8e-4_staggered/}\textquotedbl{}}
\cl{os.makedirs(output\_folder, exist\_ok=True)}
\cl{Lx = 6.283185307179586}
\cl{Ly = 6.283185307179586}
\cl{nx = 256}
\cl{\elide{257 further lines not shown}}
\end{logblock}
\inforow[Amb]{Approval}{Full script saved at: \path{agent_runs/KolmogorovBox_n256_nu8e-4_staggered/simulation.py}}
\end{agentmsgc}

\begin{agentmsgc}
Do you approve this simulation script?\par\vspace{4pt}
\btnapprove{1}\hspace{4pt}\btnfeedback{0}{Give feedback (revise)}\hspace{4pt}\btnreject{0}
\end{agentmsgc}

\begin{agentmsgc}
\stageheader[Org]{Executor}{running simulation}
\detailrow{Command: apptainer run --nv diffflowfsi.sif -g 1 \path{agent_runs/KolmogorovBox_n256_nu8e-4_staggered/simulation.py}}
\begin{logblock}{Solver Output}{63 lines}
\cl{Grid: 256 x 256, Domain: [0, 6.283185307179586] x [0, 6.283185307179586]}
\cl{Viscosity: 0.0008, Reynolds (U*L/nu): 7853.981633974483}
\cl{Chosen dt\_init (used): 0.001000 (\textless{}= dt\_CFL: 0.019634954085)}
\cl{Loop 0 completed. Runtime: 4.44 s. Time per output: 0.0444 s}
\cl{Loop 1 completed. Runtime: 3.42 s. Time per output: 0.0342 s}
\cl{All loops complete. Simulation finished.}
\end{logblock}
\okrow{Simulation finished successfully in 31.2s (exit code 0)}
\donerow{Executor}{31.2s}
\transrow{Org}{Executor}{Ros}{Post-Processor}{simulation succeeded}
\stageheader[Ros]{Post-Processor}{analysing results}
\inforow[Ros]{PostProc}{Phase A: running generic baseline analysis...}
\inforow[Ros]{PostProc}{Analyzed 2 output file(s) \textbar{} \textbar{}u\textbar{}\_max = 2.7103}
\inforow[Ros]{PostProc}{Generated 8 baseline plot(s)}
\inforow[Ros]{PostProc}{Phase B: requesting LLM custom analysis...}
\steprow{PostProc}
\inforow[Ros]{PostProc}{Custom analysis produced 14 metrics}
\inforow[Ros]{PostProc}{Running physics plausibility check...}
\steprow{PostProc}
\inforow[Ros]{PostProc}{Physics plausibility: suspicious}
\warnrow{Physics suspicious: The linear balance (alpha + nu k\^{}2 = 0.075 + 0.0008*25 = 0.095) implies a steady Fourier-mode amplitude U $\approx$ 1.5/0.095 $\approx$ 15.8, far larger than the reported u\_max (1.4 $\rightarrow$ 2.7); v is essentially zero while mean KE jumps \textasciitilde{}6$\times$ between loops. This could be an early-transient effect (very short simulated time) but also suggests checking the forcing implementation (amplitude, wavenumber), friction term, and that \elide{truncated}}
\donerow{Post-Processor}{9 plots total}
\transrow{Ros}{Post-Processor}{Ind}{Research Prep}{handing off to research agent}
\end{agentmsgc}

\begin{agentmsgc}
\begin{logblock}{Results summary}{json}
\cl{\{}
\cl{  \textquotedbl{}fluid\_loop\_1.nc.dimensions\textquotedbl{}: \textquotedbl{}\{'time': 100, 'x': 256, 'y': 256\}\textquotedbl{},}
\cl{  \textquotedbl{}fluid\_loop\_1.nc.u\_max\textquotedbl{}: \textquotedbl{}2.710273265838623\textquotedbl{},}
\cl{  \textquotedbl{}fluid\_loop\_1.nc.v\_max\textquotedbl{}: \textquotedbl{}0.007780217099934816\textquotedbl{},}
\cl{  \textquotedbl{}fluid\_loop\_1.nc.velocity\_mag\_max\textquotedbl{}: \textquotedbl{}2.7102737426757812\textquotedbl{},}
\cl{  \textquotedbl{}fluid\_loop\_1.nc.pressure\_max\textquotedbl{}: \textquotedbl{}0.016836369410157204\textquotedbl{},}
\cl{  \textquotedbl{}physics\_plausibility.verdict\textquotedbl{}: \textquotedbl{}suspicious\textquotedbl{}}
\cl{  \elide{30 further keys not shown}}
\cl{\}}
\end{logblock}
\end{agentmsgc}

\begin{agentmsgc}
\inforow[Ros]{PostProc}{Generated 9 plot(s)}
\end{agentmsgc}

\begin{agentmsgc}
ke time
\imgel{0.42\linewidth}{ke_time.png}
\end{agentmsgc}

\begin{agentmsgc}
\detailrow{\elide{8 further plots not shown}}
\end{agentmsgc}

\begin{agentmsgc}
\kvrow{Output files (4)}{\path{agent_runs/KolmogorovBox_n256_nu8e-4_staggered}}
\end{agentmsgc}

\begin{agentmsgc}
\stageheader[Ind]{Research Prep}{preparing research (combined mode)}
\inforow[Ind]{ResearchPrep}{Question: Simulate 2D forced turbulence in a doubly periodic 2*pi box on a 256x256 grid with viscosity 8e-4, driven by a Kolmogorov forcing 1.5*sin(5*y) in x plus a \elide{truncated}}
\inforow[Ind]{ResearchPrep}{Script: \path{agent_runs/KolmogorovBox_n256_nu8e-4_staggered/simulation.py}}
\inforow[Ind]{ResearchPrep}{Budget: 30 experiments}
\donerow{Research Prep}{ready}
\stageheader[Ind]{Research Initializer}{setting up research session}
\detailrow{Solver context: 7880 chars}
\steprow{Constraints}
\inforow[Ind]{ResearchInit}{Target component (session-level): forcing}
\steprow{ScoringProtocol}
\inforow[Ind]{ResearchInit}{Scoring protocol decided by the agent (question silent): metric\_averaging\_time=6, warmup\_sim\_time=26.67}
\detailrow{time scale forcing\_length\_Lf = 1.257}
\detailrow{time scale turbulent\_velocity\_u\textasciitilde{}sqrt(A*Lf) = 1.373}
\detailrow{time scale turnover\_time\_tau\_turn = Lf/u = 0.915}
\detailrow{time scale domain\_cross\_time\_Lx/u = 4.576}
\detailrow{time scale friction\_damping\_time\_1/alpha = 13.33}
\detailrow{time scale viscous\_time\_Lf\^{}2/nu = 1974}
\detailrow{Protocol reasoning: Key scales: forcing length Lf = 2$\pi$/5 $\approx$1.2566, turbulent velocity estimate u$\approx$$\surd$(A$\cdot$Lf)$\approx$1.373, giving a large-eddy turnover $\tau$$\approx$0.92; domain crossing $\approx$4.58; linear-friction damping 1/$\alpha$=13.33 (the slowest relevant process); viscous diffusion at the forcing scale is O(2$\times$10\^{}3) and not the equilibration limiter. Warm-up is set to \elide{truncated}}
\warnrow{Scoring protocol: agent left gradient\_rollout\_time undecided --- session defaults apply for those}
\inforow[Ind]{ResearchInit}{Invariants (13 --- must not change)}
\detailrow{- Lx $\rightarrow$ Domain length in x = 2$\pi$ --- fixes the physical box size (doubly-periodic 2$\pi$$\times$2$\pi$) and the wavenumber discretisation; must \elide{truncated}}
\detailrow{- Ly $\rightarrow$ Domain length in y = 2$\pi$ --- fixes the physical box size (doubly-periodic 2$\pi$$\times$2$\pi$) and the wavenumber discretisation; must \elide{truncated}}
\detailrow{\elide{11 further invariants not shown}}
\inforow[Ind]{ResearchInit}{Parameter constraints (8)}
\detailrow{- CFL/stability: the time step must satisfy the CFL constraint used by the time\_step helper (dt derived from dt\_init, cfl\_safety, grid, viscosity). Practically: \elide{truncated}}
\detailrow{\elide{7 further constraints not shown}}
\inforow[Ind]{ResearchInit}{Pre-registered mechanism categories (8): fixed\_kolmogorov\_forcing, spectral\_amplitude\_feedback\_forcing, local\_velocity\_feedback\_forcing, constant\_linear\_friction, energy\_dependent\_friction, constant\_viscosity, eddy\_viscosity\_model, spectral\_hyperviscosity}
\detailrow{Research output directory: \path{agent_runs/research_simulate_2d_forced_turbulence_in_a_doubly_periodic_grad}}
\inforow[Ind]{ResearchInit}{Scoring protocol: M=27 warm-up windows, K=6 metric windows,}
\inforow[Ind]{ResearchInit}{Base script extended to num\_loops\textgreater{}=60 (warmup 27 + metric 6 + 27 margin for the baseline check) so baseline and forward-probe runs reach stationary scoring windows.}
\donerow{Research Initializer}{starting: Simulate 2D forced turbulence in a doubly periodic 2*pi box \elide{truncated}}
\transrow{Ind}{Initializer}{Org}{Executor}{running baseline}
\end{agentmsgc}

\begin{agentmsgc}
\stageheader[Org]{Executor}{running simulation}
\detailrow{Command: apptainer run --nv diffflowfsi.sif -g 1 \path{agent_runs/research_simulate_2d_forced_turbulence_in_a_doubly_periodic_grad/exp_000/simulation.py}}
\begin{logblock}{Solver Output}{585 lines}
\cl{Grid: 256 x 256, Domain: [0, 6.283185307179586] x [0, 6.283185307179586]}
\cl{Viscosity: 0.0008, Reynolds (U*L/nu): 7853.981633974483}
\cl{Chosen dt\_init (used): 0.001000 (\textless{}= dt\_CFL: 0.019634954085)}
\cl{Loop 0 completed. Runtime: 4.46 s. Time per output: 0.0446 s}
\cl{    \elide{58 loops not shown}}
\cl{Loop 59 completed. Runtime: 0.47 s. Time per output: 0.0047 s}
\cl{All loops complete. Simulation finished.}
\end{logblock}
\okrow{Simulation finished successfully in 74.3s (exit code 0)}
\donerow{Executor}{74.3s}
\stageheader[Grn]{Research Analyzer}{analyzing Baseline}
\steprow{Analyzer}
\donerow{Research Analyzer}{Baseline: status=completed, score=36.06, metrics=5 values}
\kvrow{Analysis}{Baseline run completed with score 36.0587 (spectrum\_mse\_log = 0.36059); this is the best result so far (only experiment) and the run appears stable/converged (n\_converged\_files = 50, k range 2--80). To improve the match to the reference spectrum, explore modifications to the dissipation: tune the linear friction around the present 0.075 (e.g., try a sweep roughly 0.02--0.2) and/or add \elide{truncated}}
\okrow{Exp 0: score=36.06 (baseline)}
\detailrow{1 completed \textbar{} Best: score=36.06 (Exp \#0)}
\inforow[Grn]{Analyzer}{JAX metric validated: score = 7318.56}
\inforow[Grn]{Analyzer}{JAX metric saved: \path{agent_runs/research_simulate_2d_forced_turbulence_in_a_doubly_periodic_grad/_jax_metric.py} (4938 chars)}
\inforow[Grn]{Analyzer}{Baseline rescored from saved windows [27, 28, 29, 30, 31, 32] (K=6): snippet=36.0587 $\rightarrow$ 39.3901}
\inforow[Grn]{Analyzer}{Scoring protocol check: baseline per-window score stationary from window 7 (warm-up M=27, K=6)}
\transrow{Grn}{Analyzer}{Amb}{Research Approval}{baseline completed}
\end{agentmsgc}

\begin{agentmsgc}
\stageheader[Amb]{Research Approval}{awaiting user review}
\renewcommand{\planaccent}{uiBlue}
\begin{plancard}{Research Plan}
\plansec{Research}
\planrow{Question}{Simulate 2D forced turbulence in a doubly periodic 2*pi box on a 256x256 grid with viscosity 8e-4, driven by a Kolmogorov forcing 1.5*sin(5*y) in x plus a linear friction of 0.075. Then improve the friction/dissipation model so that the energy spectrum matches the reference spectrum saved at reference\_spectrum.npz (keys 'k' and 'E'). Only change the forcing and dissipation model; no neural networks.}
\plansec{Baseline}
\planrow{Status}{completed}
\planrow{spectrum\_mse\_log}{0.3605866797516565}
\planrow{score}{39.39012146}
\planrow{k\_min}{2.0}
\planrow{k\_max}{80.0}
\planrow{n\_converged\_files}{50}
\planrow{baseline\_rescore\_windows\_used}{[27, 28, 29, 30, 31, 32]}
\planrow{baseline\_per\_loop\_scores}{[7318.455566, 4376.308594, 194 \elide{truncated}  40.4690361, 39.86376953]}
\planrow{baseline\_score\_stationary\_from\_window}{7}
\planrow{baseline\_score\_stationarity\_converged}{True}
\planrow{warmup\_windows\_sufficient}{True}
\plansec{Analysis}
\planline{Baseline run completed with score 36.0587 (spectrum\_mse\_log = 0.36059); this is the best result so far (only experiment) and the run appears stable/converged (n\_converged\_files = 50, k range 2--80). To improve the match to the reference spectrum, explore modifications to the dissipation: tune the linear friction around the present 0.075 (e.g., try a sweep roughly 0.02--0.2) and/or add scale-selective small-scale damping (hyperviscous-like dissipation) to remove excess high‑k energy and reduce the spectrum MSE.}
\end{plancard}
\end{agentmsgc}

\begin{agentmsgc}
Proceed with auto-research experiments?\par\vspace{4pt}
\btnapprove{1}\hspace{4pt}\btnfeedback{0}{Give feedback (revise)}\hspace{4pt}\btnreject{0}
\end{agentmsgc}

\begin{omitted}
{\bfseries Experiment 1 not shown}
\end{omitted}

\begin{agentmsg}
\stageheader[Vio]{Research Proposer}{planning next step (28 experiments remaining)}
\steprow{Proposer}
\donerow{Research Proposer}{experiment: Optimizing the spectral hyperviscosity amplitude (in log-space) and its onset cutoff will \elide{truncated}}
\kvrow{Hypothesis}{Optimizing the spectral hyperviscosity amplitude (in log-space) and its onset cutoff will preferentially remove excess high-k energy and reduce the spectrum\_mse\_log relative to the current probe, improving the match to the reference spectrum.}
\kvrow{Parameter changes}{\{\textquotedbl{}\_log\_hypervisc\_strength\textquotedbl{}: -5.0, \textquotedbl{}\_hypervisc\_k\_cut\textquotedbl{}: 64.0\}}
\kvrow{Gradient request}{optimize $\cdot$ 10 epochs $\cdot$ lr 0.2 $\cdot$ clip\_by\_global\_norm, adam}
\transrow{Vio}{Proposer}{Blu}{Modifier}{next experiment}
\stageheader[Blu]{Research Modifier}{creating experiment 2}
\steprow{Modifier}
\donerow{Research Modifier}{experiment 2 script saved}
\stageheader[Cya]{Research Validator}{checking modified script}
\warnrow{Preflight smoke test failed: partial\_disconnect --- validation rejected before executor}
\warnrow{Validation found 1 error(s) - Preflight smoke test (partial\_disconnect): Preflight smoke test: gradient is negligible on these trainable fields: ['leaf\_1'], relative to the loss (metric$\approx$23.4) --- they have almost no leverage to reduce it. Act on WHICH layer the \elide{truncated}}
\transrow{Cya}{Validator}{Blu}{Modifier}{validation failed --- retry 1/3}
\stageheader[Blu]{Research Modifier}{fixing experiment 2 (retry 1)}
\steprow{Modifier-Retry}
\inforow[Blu]{Modifier-Retry}{Retry produced complete-script output (no patches) --- accepted}
\donerow{Research Modifier}{retry fix saved for experiment 2}
\stageheader[Cya]{Research Validator}{checking modified script}
\detailrow{Preflight passed: Preflight smoke test passed.}
\donerow{Research Validator}{all checks passed}
\transrow{Cya}{Validator}{Org}{Executor}{validation passed}
\stageheader[Org]{Executor}{running simulation}
\begin{logblock}{Solver Output}{71 lines}
\cl{  Epoch 0/10: loss=41.0116, \textbar{}\textbar{}grad\textbar{}\textbar{}=3.299e+00 (42.3s)}
\cl{  Epoch 1/10: loss=53.7557, \textbar{}\textbar{}grad\textbar{}\textbar{}=5.333e+00 (10.2s)}
\cl{    \elide{6 epochs not shown}}
\cl{  Epoch 8/10: loss=14.3242, \textbar{}\textbar{}grad\textbar{}\textbar{}=2.688e+01 (9.5s)}
\cl{  Epoch 9/10: loss=11.3467, \textbar{}\textbar{}grad\textbar{}\textbar{}=5.395e+00 (9.3s)}
\cl{OPTIMIZATION\_STATUS: completed}
\cl{OPTIMIZATION\_INITIAL\_LOSS: 41.0116}
\cl{OPTIMIZATION\_BEST\_LOSS: 11.3467}
\end{logblock}
\okrow{Simulation finished successfully in 187.6s (exit code 0)}
\donerow{Executor}{187.6s}
\stageheader[Grn]{Research Analyzer}{analyzing Experiment 2}
\steprow{Analyzer}
\detailrow{Gradients: $\partial$/$\partial$log\_strength=-5.303, $\partial$/$\partial$k\_cut=0.994}
\detailrow{Optimization: 10 epochs, loss 41.01 $\rightarrow$ 11.35}
\donerow{Research Analyzer}{Experiment 2: status=completed, score=8.179, metrics=3 values}
\kvrow{Analysis}{Score = 8.179251671 (lower is better), a large improvement over the prior best 39.0203 --- this is now the best-so-far. The optimizer moved log\_strength from -5 $\rightarrow$ -3.200 (S $\approx$ 10\^{}(-3.2) $\approx$ 6.3$\times$10\^{}-4) with k\_cut $\approx$ 62.19; gradients ($\partial$score/$\partial$log\_strength $\approx$ -5.30, $\partial$score/$\partial$k\_cut $\approx$ +0.99) indicate increasing the hypervisc strength would further reduce the score while increasing k\_cut slightly worsens it. \elide{truncated}}
\okrow{Exp 2: score=8.179 (\_log\_hypervisc\_strength=-5.0, \_hypervisc\_k\_cut=64.0) \textbar{} grad: $\partial$/$\partial$log\_strength=-5.3, $\partial$/$\partial$k\_cut=0.994}
\detailrow{3 completed \textbar{} Best: score=8.179 (Exp \#2)}
\inforow[Grn]{Analyzer}{Base script updated to exp 2 (score 8.18 \textless{} prev best 39.02)}
\transrow{Grn}{Analyzer}{Vio}{Proposer}{experiment recorded}
\progressbar{2}{30}
\end{agentmsg}

\begin{omitted}
{\bfseries Experiments 3--16 not shown}
\end{omitted}

\begin{agentmsg}
\stageheader[Vio]{Research Proposer}{planning next step (13 experiments remaining)}
\steprow{Proposer}
\donerow{Research Proposer}{experiment: A small, band-limited spectral amplitude feedback (gain=0.02, bandwidth$\approx$1.0, clip $\pm$5\%) \elide{truncated}}
\kvrow{Hypothesis}{A small, band-limited spectral amplitude feedback (gain=0.02, bandwidth$\approx$1.0, clip $\pm$5\%) that nudges injection at k=5 toward the reference band energy (ratio-based correction) will reduce the spectrum misfit near the forcing wavenumber and lower the overall spectrum\_mse\_log relative to exp\_16, while remaining numerically stable.}
\kvrow{Parameter changes}{\{\textquotedbl{}kol\_amplitude\textquotedbl{}: 1.396, \textquotedbl{}energy\_fric\_gain\textquotedbl{}: 0.5829, \textquotedbl{}\_log\_hypervisc\_strength\textquotedbl{}: -3.405, \textquotedbl{}\_hypervisc\_k\_cut\textquotedbl{}: 61.99, \textquotedbl{}\_feedback\_gain\textquotedbl{}: 0.02, \textquotedbl{}\_feedback\_k\_center\textquotedbl{}: 5.0, \textquotedbl{}\_feedback\_k\_width\textquotedbl{}: 1.0, \textquotedbl{}\_feedback\_clip\_max\textquotedbl{}: 0.05, \textquotedbl{}\_feedback\_mode\textquotedbl{}: \textquotedbl{}ratio\_to\_ref\textquotedbl{}, \textquotedbl{}\_feedback\_ref\_path\textquotedbl{}: \textquotedbl{}reference\_spectrum.npz\textquotedbl{}\}}
\transrow{Vio}{Proposer}{Blu}{Modifier}{next experiment}
\stageheader[Blu]{Research Modifier}{creating experiment 17}
\steprow{Modifier}
\detailrow{Applied 2 patch(es) successfully}
\donerow{Research Modifier}{experiment 17 script saved}
\stageheader[Cya]{Research Validator}{checking modified script}
\warnrow{Validation found 3 error(s) - `\_feedback\_mode` from `parameter\_changes` is bound at module level but never read anywhere in the script --- the recorded parameter has no effect on this run. Read it inside the closure/solver setup it parameterizes, or drop it \elide{truncated}}
\transrow{Cya}{Validator}{Blu}{Modifier}{validation failed --- retry 1/3}
\stageheader[Blu]{Research Modifier}{fixing experiment 17 (retry 1)}
\steprow{Modifier-Retry}
\donerow{Research Modifier}{retry fix saved for experiment 17}
\stageheader[Cya]{Research Validator}{checking modified script}
\detailrow{Preflight passed: forward-only script reached its first saved window}
\donerow{Research Validator}{all checks passed}
\transrow{Cya}{Validator}{Org}{Executor}{validation passed}
\stageheader[Org]{Executor}{running simulation}
\begin{logblock}{Solver Output}{123 lines}
\cl{Traceback (most recent call last):}
\cl{    \elide{63 traceback lines not shown}}
\cl{TypeError: '\_TrainableParams' object is not callable}
\end{logblock}
\errrow{Simulation failed (exit code 1) after 22.9s}
\donerow{Executor}{FAILED}
\stageheader[Grn]{Research Analyzer}{analyzing Experiment 17}
\steprow{Analyzer}
\donerow{Research Analyzer}{Experiment 17: status=crashed, score=inf, metrics=0 values}
\kvrow{Analysis}{The run crashed (score = inf; no metrics recorded) so this is not better than the best-so-far score 2.79975. The traceback shows a Python TypeError in friction\_forcing (simulation.py:153): a '\_TrainableParams' object was placed where a callable forcing was expected (model(state)), so this is an implementation/runtime bug (wrong object inserted into the forcings list), not a numerical instability \elide{truncated}}
\warnrow{Exp 17: crashed (kol\_amplitude=1.396, energy\_fric\_gain=0.5829, \_log\_hypervisc\_strength=-3.405)}
\detailrow{16 completed, 2 failed \textbar{} Best: score=2.8 (Exp \#16)}
\transrow{Grn}{Analyzer}{Vio}{Proposer}{experiment recorded}
\progressbar{17}{30}
\end{agentmsg}

\begin{omitted}
{\bfseries Experiments 18--26 not shown}
\end{omitted}

\begin{agentmsg}
\stageheader[Vio]{Research Proposer}{planning next step (3 experiments remaining)}
\steprow{Proposer}
\donerow{Research Proposer}{experiment: Floating and optimizing the three scalar fields of exp\_012's LocalEnergyDamping \elide{truncated}}
\kvrow{Hypothesis}{Floating and optimizing the three scalar fields of exp\_012's LocalEnergyDamping (local\_feedback\_gain, local\_feedback\_sigma, local\_feedback\_E\_ref) will allow the optimizer to find a locally better damping configuration that reduces spectrum\_mse\_log below the current best ($\approx$2.322) or else will show the mechanism cannot compete when given a fair scalar tuning.}
\kvrow{Parameter changes}{\{\textquotedbl{}\_local\_feedback\_gain\textquotedbl{}: 0.02, \textquotedbl{}\_local\_feedback\_sigma\textquotedbl{}: 0.5, \textquotedbl{}\_local\_feedback\_E\_ref\textquotedbl{}: 0.14\}}
\kvrow{Gradient request}{optimize $\cdot$ 12 epochs $\cdot$ lr 0.01 $\cdot$ clip\_by\_global\_norm, adam}
\transrow{Vio}{Proposer}{Blu}{Modifier}{next experiment}
\stageheader[Blu]{Research Modifier}{creating experiment 27}
\steprow{Modifier}
\donerow{Research Modifier}{experiment 27 script saved}
\stageheader[Cya]{Research Validator}{checking modified script}
\warnrow{Preflight smoke test failed: jit\_crash --- validation rejected before executor}
\warnrow{Validation found 1 error(s) - Preflight smoke test (jit\_crash): Preflight smoke test crashed before Step 5 completed: Traceback (most recent call last): ... [17 solver-internal frame(s) elided] ... File \elide{truncated}}
\transrow{Cya}{Validator}{Blu}{Modifier}{validation failed --- retry 1/3}
\stageheader[Blu]{Research Modifier}{fixing experiment 27 (retry 1)}
\steprow{Modifier-Retry}
\donerow{Research Modifier}{retry fix saved for experiment 27}
\stageheader[Cya]{Research Validator}{checking modified script}
\detailrow{Preflight passed: Preflight smoke test passed.}
\donerow{Research Validator}{all checks passed}
\transrow{Cya}{Validator}{Org}{Executor}{validation passed}
\stageheader[Org]{Executor}{running simulation}
\begin{logblock}{Solver Output}{86 lines}
\cl{  Epoch 0/12: loss=6.76876, \textbar{}\textbar{}grad\textbar{}\textbar{}=3.173e+02 (69.2s)}
\cl{  Epoch 1/12: loss=3.77255, \textbar{}\textbar{}grad\textbar{}\textbar{}=1.499e+02 (32.3s)}
\cl{  Epoch 2/12: loss=2.65245, \textbar{}\textbar{}grad\textbar{}\textbar{}=0.000e+00 (13.1s)}
\cl{  Epoch 3/12: loss=3.52277, \textbar{}\textbar{}grad\textbar{}\textbar{}=0.000e+00 (12.8s)}
\cl{  Epoch 4/12: loss=2.9589, \textbar{}\textbar{}grad\textbar{}\textbar{}=0.000e+00 (12.7s)}
\cl{  Epoch 5/12: loss=2.15851, \textbar{}\textbar{}grad\textbar{}\textbar{}=0.000e+00 (12.6s)}
\cl{OPTIMIZATION\_STATUS: completed}
\cl{OPTIMIZATION\_INITIAL\_LOSS: 6.76876}
\cl{OPTIMIZATION\_BEST\_LOSS: 2.15851}
\end{logblock}
\okrow{Simulation finished successfully in 258.0s (exit code 0)}
\donerow{Executor}{258.0s}
\stageheader[Grn]{Research Analyzer}{analyzing Experiment 27}
\steprow{Analyzer}
\warnrow{Gradient: saturated The best model's gradients are exactly 0.0 on every parameter, but the epoch-0 preflight gradients were nonzero: the wrapper is wired correctly and the best parameters sit in a locally flat region (typically a clamp/floor/cap bound). The \elide{truncated}}
\inforow[Grn]{Analyzer}{Gradient diagnosis: Every trainable-parameter gradient is exactly 0.0 on the completed run: the loss is locally flat in}
\detailrow{Optimization: 6 epochs, loss 6.769 $\rightarrow$ 2.159}
\inforow[Grn]{Analyzer}{Optimization diagnosis: Optimization measured NOTHING: every epoch's loss stayed within the evaluation-noise band (2$\sigma$ = 6.55) around the start,}
\donerow{Research Analyzer}{Experiment 27: status=completed, score=2.103, metrics=3 values}
\kvrow{Analysis}{Training dynamics (concrete) - When did loss stabilize? The loss fell rapidly early and essentially settled by epoch 5 (best/final loss = 2.15851, best epoch = 5/5). The final reported score (2.102755547) comes from the completed evaluation and the epoch-to-epoch loss was not still descending at the end --- it plateaued. - Shape diagnosis from the trajectory: the loss began high (6.76876), dropped \elide{truncated}}
\okrow{Exp 27: score=2.103 (\_local\_feedback\_gain=0.02, \_local\_feedback\_sigma=0.5, \_local\_feedback\_E\_ref=0.14) \textbar{} grad: saturated (params at guard/clamp bound)}
\detailrow{25 completed, 3 failed \textbar{} Best: score=2.103 (Exp \#27)}
\inforow[Grn]{Analyzer}{Base script updated to exp 27 (score 2.10 \textless{} prev best 2.32)}
\transrow{Grn}{Analyzer}{Vio}{Proposer}{experiment recorded}
\progressbar{27}{30}
\end{agentmsg}

\begin{omitted}
{\bfseries Experiments 28--29 not shown}
\end{omitted}

\begin{agentmsg}
\stageheader[Mag]{Research Synthesizer}{generating final report}
\steprow{Synthesizer}
\donerow{Research Synthesizer}{research complete}
\renewcommand{\planaccent}{uiBlue}
\begin{plancard}{Research Summary}
\plansec{1. Research Question}
\planline{Simulate 2D forced turbulence in a doubly periodic 2$\pi$ box on a 256$\times$256 grid with viscosity $\nu$ = 8e-4, Kolmogorov forcing amplitude 1.5$\cdot$sin(5 y) in x and linear friction $\alpha$ = 0.075. Modify only the forcing and dissipation models (no NNs) to make the simulated energy spectrum match the reference spectrum at reference\_spectrum.npz (keys 'k' and 'E'). Which forcing/dissipation modifications achieve the best spectrum match (minimize spectrum\_mse\_log)?}
\plansec{2. Methodology}
\planline{Approach: Iterative, automated probes across mechanism families (spectral hyperviscosity, energy-dependent friction, local/ spectral feedback forcings, Smagorinsky eddy-viscosity), mixing forward-only probes, gradient-only \elide{truncated}}
\planline{Parameters varied (examples): kol\_amplitude, linear friction (alpha), spectral hypervisc strength (log scale) and k\_cut, energy-dependent friction gain/E\_ref, local velocity feedback gains/sigma/E\_ref, Smagorinsky C and filter \elide{truncated}}
\planline{Baseline configuration: Experiment 0 (unmodified) with the stated $\nu$, forcing 1.5$\cdot$sin(5 y), $\alpha$=0.075.}
\planline{Experiment tally: 30 experiments: 27 completed, 2 invalid, 1 crashed (statuses listed below).}
\plansec{3. Experiment Results}
\planline{Summary table (ID, Score, Status, Changes)}
\planline{\elide{table of 30 experiments not shown}}
\end{plancard}
\end{agentmsg}

\begin{agentmsgc}
Comparison
\imgel{0.88\linewidth}{comparison.png}
\detailrow{Baseline (exp 0), score 39.39 vs Best (exp 27), score 2.103, each over the saved windows its score was computed on, with the metric's reference where it has one.}
\end{agentmsgc}

\begin{agentmsgc}
Score Progress
\imgel{0.5\linewidth}{progress.png}
\end{agentmsgc}

\begin{agentmsgc}
\kvrow{Output files}{\path{agent_runs/research_simulate_2d_forced_turbulence_in_a_doubly_periodic_grad}}
\kvrow{Best script}{\path{agent_runs/research_simulate_2d_forced_turbulence_in_a_doubly_periodic_grad/best_script.py}}
\par\vspace{2pt}Workflow complete (budget exhausted).
\end{agentmsgc}
\chatframeoff

\bibliographystyle{elsarticle-num-names}
\bibliography{main}

\end{document}